\documentclass[useAMS,usenatbib]{mn2e}
\usepackage{graphicx}
\title[Ionized outflows in SDSS type 2 quasars]{Ionized outflows in  SDSS type 2 quasars at $z\sim$0.3-0.6\thanks{Based on observations carried out at the
European Southern Observatory (Paranal, Chile) with FORS2 on VLT-UT1 (programmes
081.B-0129 and 083.B-0381.}}
\author[Villar-Mart\'\i n et al.]{M. Villar-Mart\'\i n$^{1,2}$, A. Humphrey$^3$, R. Gonz\'alez Delgado$^1$, L. Colina$^2$,
S. Arribas$^2$ \\
$^{1}$Instituto de Astrof\'\i sica de Andaluc\'\i a (CSIC), Glorieta de la Astronom\'\i a s/n, 18008 Granada, Spain. villarmm@cab.inta-csic.es \\
$^2$Centro de Astrobiolog\'\i a (INTA-CSIC), Carretera de Ajalvir, km 4, 28850 Torrej\'on de Ardoz, Madrid, Spain\\
$^{3}$Instituto Nacional de Astrof\'\i sica, Optica y Electr\'onica (INAOE), Aptdo. Postal 51 y 216, 72000 Puebla, M\'exico}

\begin{document}

\date{} 

\pagerange{\pageref{firstpage}--\pageref{lastpage}} \pubyear{2002}

\maketitle

\begin{abstract}

We have analyzed the spatially integrated kinematic properties of the ionized gas within the inner $r\la$few kpc in  13 optically selected SDSS type 2 quasars at $z\sim$0.3-0.6, using the [OIII]$\lambda\lambda$4959,5007 lines.
The line profiles show a significant asymmetry in 11 objects. There is a clear preference for blue asymmetries, which are found in 9/13 quasars at 10\% intensity level.  In coherence with studies on other types of active and non active galaxies, we propose that the asymmetries are produced by outflows where differential dust extinction is at work.

This scenario is favoured by other results we find:  in addition to quiescent ambient gas, whose kinematic properties are consistent with gravitational motions, we have discovered highly perturbed gas  in all objects. This gas emits very broad lines ($R=\frac{FWHM[OIII]}{FWHM_ {stars}}\ga$2). While the quiescent gas shows small or null velocity shifts relative to the systemic velocity, the highly perturbed gas trends to show larger shifts which, moreover, are blueshifts in general.
Within a given object, the most perturbed gas  trends to have the largest blueshift as well.
 All together support that the perturbed gas, which is responsible for the blue asymmetries  of the line profiles, is outflowing. The outflowing gas is located within the quasar ionization cones, in the narrow line region.
  
The relative contribution of the outflowing gas  
	to the total [OIII] line flux varies from object to object in the range $\sim$10-70\%.  An anticorrelation is found such that, the more perturbed the outflowing gas is, the lower its relative contribution is  to the total [OIII] flux . 
This suggests that outflows with more perturbed kinematics  involve a smaller fraction of the total
 mass of ionized gas.

Although some bias affects the sample, we argue that ionized gas outflows are a common phenomenon in optically selected type 2 quasars at 0.3$\la z \la$0.6.

\end{abstract}


\label{firstpage}

\section{Introduction}

Evidence for an intimate connection between supermassive black hole (SMBH) growth and
 evolution of galaxies           is
nowadays compelling. Not only have SMBHs  been found in many
galaxies with a bulge component, but tight correlations exist between the black hole mass and some bulge
properties, such as the stellar mass and velocity dispersion $\sigma$ (Ferrarese \& Merrit \citeyear{fer00}, Gebhardt et al. \citeyear{geb00}). The origin of this relation is still an open question, but quasar induced outflows might play a critical role. Hydrodinamycal simulations show that the energy input from quasars can regulate the growth and activity of black holes and their host galaxies (di Matteo, Springel \& Hernquist \citeyear{dimatteo05}). Such models show that a merger between two galaxies with SMBH leads to strong inflow that feeds gas to the SMBH activating the quasar phase.
The energy released by the strong  outflows associated with major phases of accretion  expels enough gas to quench both star formation and further black hole growth. This determines the lifetime of the quasar phase and explains the relationship between the black hole mass  and the stellar velocity dispersion.  

This scenario, although vey attractive, still lacks compelling observational support. The presence of outflows in different classes of active galaxies has been known for several decades. However, it is still not clear whether they can generate important  feedback effects with a notable impact on the systems evolution. The limitations are several, such as the uncertainty on where the outflows originate and their geometry, or the fact that several gas phases (molecular, neutral, ionized) can be involved. These are often observable at different spectral ranges and spatial scales and therefore require different observing techniques.

The presence of outflows of ionized gas  in different types of active galaxies is well stablished.   They are evident, for instance, as absorption lines in the UV and X-ray spectra, blueshifted relative to the systemic velocity of the host galaxy. Outflowing velocities of up to $\sim$2000 km s$^{-1}$ have been measured
(see Crenshaw, Kraemer \& George \citeyear{cren99} for a review).

Outflows of ionized gas have also been discovered  when characterizing the gas
  kinematics  using the spectral properties of  a diversity of  forbidden lines in  the infared (e.g. Spoon \& Holt \citeyear{spoo10}) and the optical.
 The [OIII]$\lambda\lambda$4959,5007 lines have been used more frequently in optical studies.   These are often the strongest optical lines and have the advantage of being cleanly isolated from other emission and absorption features. In addition, underlying contamination by line emission from the broad line region (BLR) is not expected, as is often the case for permitted lines such as H$\alpha$ and H$\beta$.

The predominance of blue assymetries on the [OIII] line profiles,  their frequent blueshift  relative to different indicators of the systemic redshift   and the kinematic substructure  of the lines indicate that the  narrow line region (NLR) gas is not a stable, quiescent gas reservoir that follows the bulge gravitational potential. This dynamic behaviour has often been interpreted in terms of outflows where differential reddening is present: i.e. at least part of the NLR gas is moving radially outwards and  dust reddening is more severe  for the receding gas. This applies to active galaxies in general (e.g. Greene \& Ho \citeyear{greene05}) and specific AGN types:  Seyferts 1 and 2s (Whittle et al. \citeyear{whi88}, Colbert et al. \citeyear{col96}, Crenshaw et al. \citeyear{cren10}, Heckman et al. \citeyear{heck81}, Veilleux \citeyear{vei91}), Narrow Line Seyfert 1 galaxies (Bian et al. \citeyear{bian05}, Komossa et al. \citeyear{kom08}), radio loud and radio quiet type 1 quasars (Leipski \& Bennert \citeyear{leip06}, Heckman et al. \citeyear{heck84},  Boroson \citeyear{bor05}) and narrow line radio galaxies at different redshifts   (Humphrey et al. \citeyear{hum06}, Heckman et al. \citeyear{heck81}, Holt, Tadhunter \& Morganti \citeyear{holt08}).  \cite{heck81} found that blue assymmetries are in general not found in steep spectrum (extended) radio sources, but preferentially in radio quiet objects and flat spectrum (compact) radio sources.

The origin of the outflows has also been subject of study. Correlations between the radio properties and the kinematic signatures of the outflows have  suggested  that the radio jets/lobes can play an important role to trigger such outflows  both in radio loud and radio quiet objects (e.g. Humphrey et al. \citeyear{hum06}, Leipski \& Bennert \citeyear{leip06}, Heckman et al. \citeyear{heck84}, Whittle et al. \citeyear{whi88},  etc.).  The dynamics of the ionized gas, which is subject to a partial or full radiation field from the central continuum source, may be affected by the radiation force as well  (e.g. Boroson \citeyear{bor05}). 
Finally, galactic superwinds may also exist. Star forming systems at low and high redshift are known to drive galactic winds with observable signatures in the wings of the emission  lines (e.g. Shapiro et al. \citeyear{shap09}).

Type 2 (obscured) quasars have been discovered in large quantities only in recent years (e.g. Zakamska et al. \citeyear{zak03}, Mart\'\i nez-Sansigre et al. \citeyear{san05}). From the observational point of view, these are the luminous counterparts of Seyfert 2 galaxies.
 In particular, Zakamska et al.
 identified in 2003 several hundred objects in the redshift  range
0.3$\le z \le$0.8 in the Sloan Digital Sky Survey (SDSS)  with the high ionization
narrow emission line spectra characteristic of type 2 AGNs and  narrow
line luminosities typical of type 1 quasars. 
 Their far-infrared luminosities place
them among the most luminous quasars at similar $z$. Several studies suggest that they  have high star-forming luminosities (e.g. Zakamska et al. \citeyear{zak08},  Hiner et al. \citeyear{hin09}).
They  show a wide range of X-ray luminosities and obscuring column densities
(Ptak  et al. \citeyear{ptak06}; Zakamska  et al. \citeyear{zak04}). The host galaxies are most frequently ellipticals with irregular morphologies  (Zakamska et al. \citeyear{zak06}), and
the nuclear optical emission is highly polarized in some cases (Zakamska et al. \citeyear{zak05}).  15\%$\pm$5\% qualify as radio loud  (Zakamska et al. \citeyear{zak04},  Lal \& Ho \citeyear{lal10}).
  
Because of the recent discovery, little is known about the possible existence of outflows in type 2 quasars.
 \cite{green11}  analysed the kinematics of the {\it extended} ionized gas in a sample of 15 luminous SDSS type 2 quasars at $z<$0.5 in spatial scales of $\la$10 kpc.  They found kinematically disturbed gas accross the host galaxies and proposed  that AGN driven outflows are stirring up the gas on  scales of kpcs.   

We investigate in this paper the presence of ionized gas outflows in  a sample of 13 optically selected type 2 quasars at 0.3$\la z\la$0.6.
The observations and data analysis techniques are described in $\delta$2.   The results are presented in $\delta$3 and discussed in $\delta$4.  The results and conclusions  are summarized in $\delta$5.

 We assume
$\Omega_{\Lambda}$=0.7, $\Omega_M$=0.3, H$_0$=71 km s$^{-1}$ Mpc$^{-1}$.

\section{Observations and analysis}

The data  were obtained with the Focal Reducer and Low Dispersion Spectrograph (FORS2) for the Very Large Telescope (VLT) installed on UT1
(Appenzeller et al. \citeyear{appen98}). 
The observations were performed in two different runs: 8 and 9th Sept 2008  (4 objects) and 17 and 18 April 2009 (9 objects). 
All spectra  were obtained  with the 600RI+19 grism and the   GG435+81 order sorting filter. The useful spectral range was $\sim$5030-8250 \AA\ in the
2008 run and $\sim$5300-8600 \AA\ in the 2009 run, so that in all cases at least the H$\beta$ and  [OIII]$\lambda\lambda$4959,5007 lines were within the observed spectral range. The pixel scales are 0.25$\arcsec$ pixel$^{-1}$ and 0.83
\AA\ pix$^{-1}$ in the spatial and spectral directions respectively. The spectral resolution, as measured from the sky emission lines, was  
  FWHM=7.2$\pm$0.2 \AA\ and 5.4$\pm$0.2 \AA\  for the 2008 and 2009 runs respectively. The slit width was 1.3$\arcsec$ in 2008 and
 1.0$\arcsec$ in 2009. The observations, data reduction and object sample are described in detail in Villar-Mart\'\i n et al. (\citeyear{vil11},\citeyear{vil10}).

1-dim spectra were extracted  from spatial apertures centered on the quasars spatial continuum centroid. The  sizes were selected to optimize the S/N of the nuclear emission lines, in particular for the detection of faint broad wings. In general, the aperture sizes are $\sim$1.5-2$\arcsec$, corresponding to physical sizes $\sim$7-12 kpc 
depending on the object.  
Type 2 quasars are often associated with extended emission line regions (EELR)  which can extend sometimes for tens of kpc (e.g. Villar-Mart\'\i n et al. \citeyear{vil11}), well beyond the size of the NLR. The apertures we have defined contain emission from both the NLR and, when existing,  the EELR, although the contribution of the NLR is likely to dominate.

The spectral profiles of  [OIII] and H$\beta$  were fitted  with the STARLINK DIPSO package. One or more Gaussian components were used depending on the quality of the fits. This was evaluated based
both  on the errors of the fits and a visual inspection of the residuals. The [OIII] lines are in all cases the strongest and they were used to constrain the expected kinematic substructure of  H$\beta$. The ratio between the [OIII]$\lambda\lambda$4959,5007 components was forced to be 1:3, in accord with the theoretical value (Osterbrock 1989). Both components were forced to have the same FWHM and in general  the separation in wavelength was also  fixed.  We found necessary to apply these constraints to all kinematic components in order to reject fits that were successful at reproducing the line shapes, but did not have physical meaning.  

To start, we attempted to fit the [OIII]  profiles with a single Gaussian function, but this failed in all cases and very large residuals remained. The number of Gaussians was progressively increased  until a good fit  was achieved. Two or  three  Gaussians at most were required in all quasars.  
The H$\beta$ line profile could be decomposed in different kinematic components only in some objects where the line had enough signal and was not severely affected by atmospheric or galactic absorption.    In those cases, it was assumed that the line consists of the same number of kinematic components as [OIII]. The FWHM and $z$ of each component were constrained from those derived from [OIII].  Only those components with a detectable flux were considered in the final fit of H$\beta$.   Flux upper limits for the non detected components were also estimated. The  FWHM  (corrected for instrumental broadening in quadrature),   the velocity shift relative to the  narrow [OIII] core (see $\delta$2.1), the flux, the flux  relative to the total line flux and the [OIII]/H$\beta$ ratio  (when possible) 
were measured for all kinematics components. 

The kinematic measurements obtained for the first 5 objects in Tables 1 and 2 
are affected by slit effects (see Villar-Mart\'\i n et al. 2011).
The errors quoted  take  the resulting uncertainties into account.

The results of the fits are shown in Table 2 and Figures 1 to 13. 
 
 \begin{figure}
\includegraphics{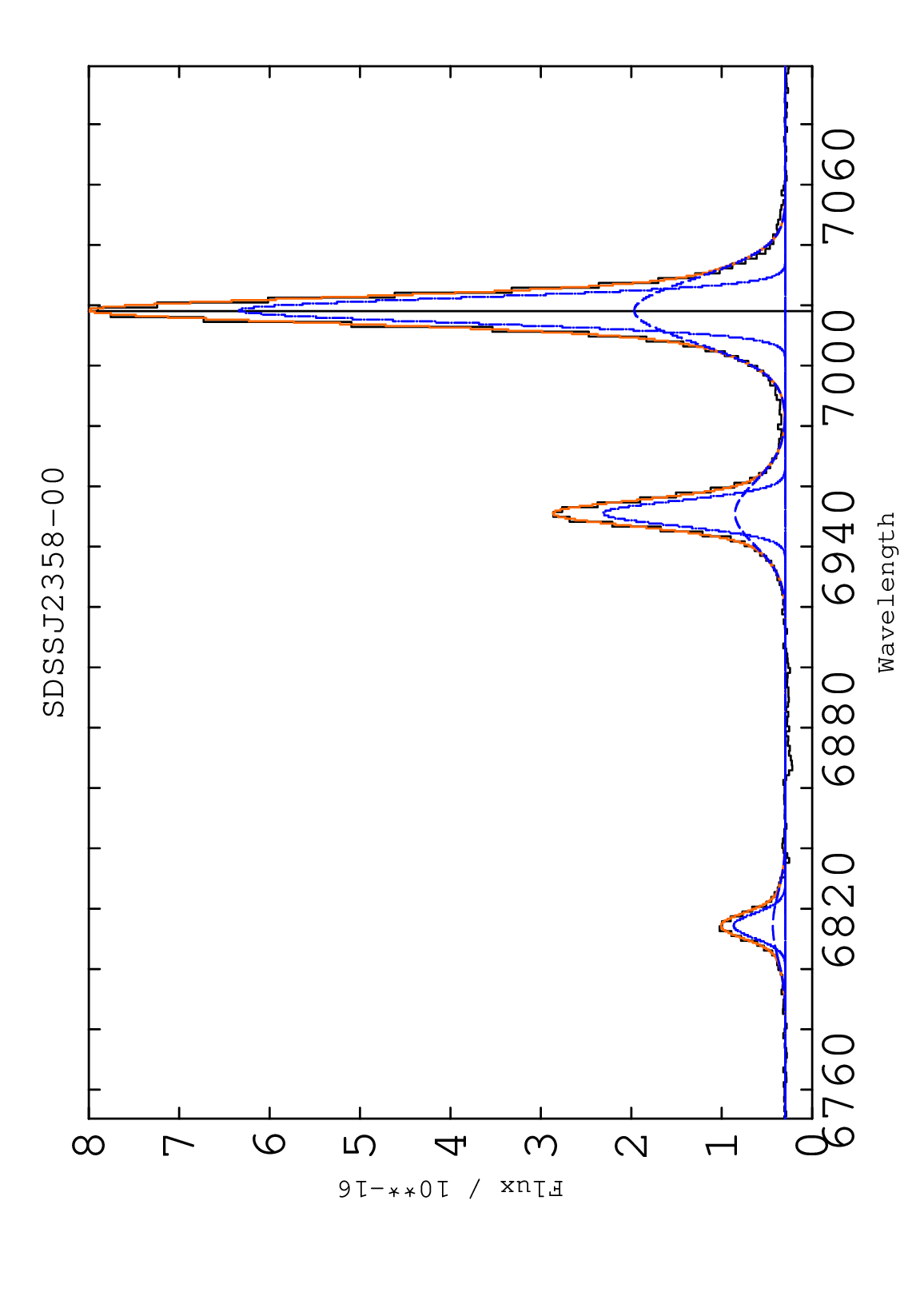}
\vspace{2.8in}
\caption{Fit to H$\beta$ and [OIII]$\lambda\lambda$4959,5007 for SDSS J2358-00. In each figure from 1 to 13 the data and the fit are shown in black and orange respectively. The same colour and line style are used for each kinematic component in all emission lines ([OIII]$\lambda\lambda$4959,5007 and, when possible, also H$\beta$).  Red  and blue are  used for the most redshifted and blueshifted components respectively.
Green is used for an intermediate redshift. When two components are identified with equal $z$ within the errors (as the example in this figure), the same colour but different line style are used for each one. The vertical black line marks the wavelength of the very narrow [OIII] peak, which we have used to caculate the velocity shifts. }
\includegraphics{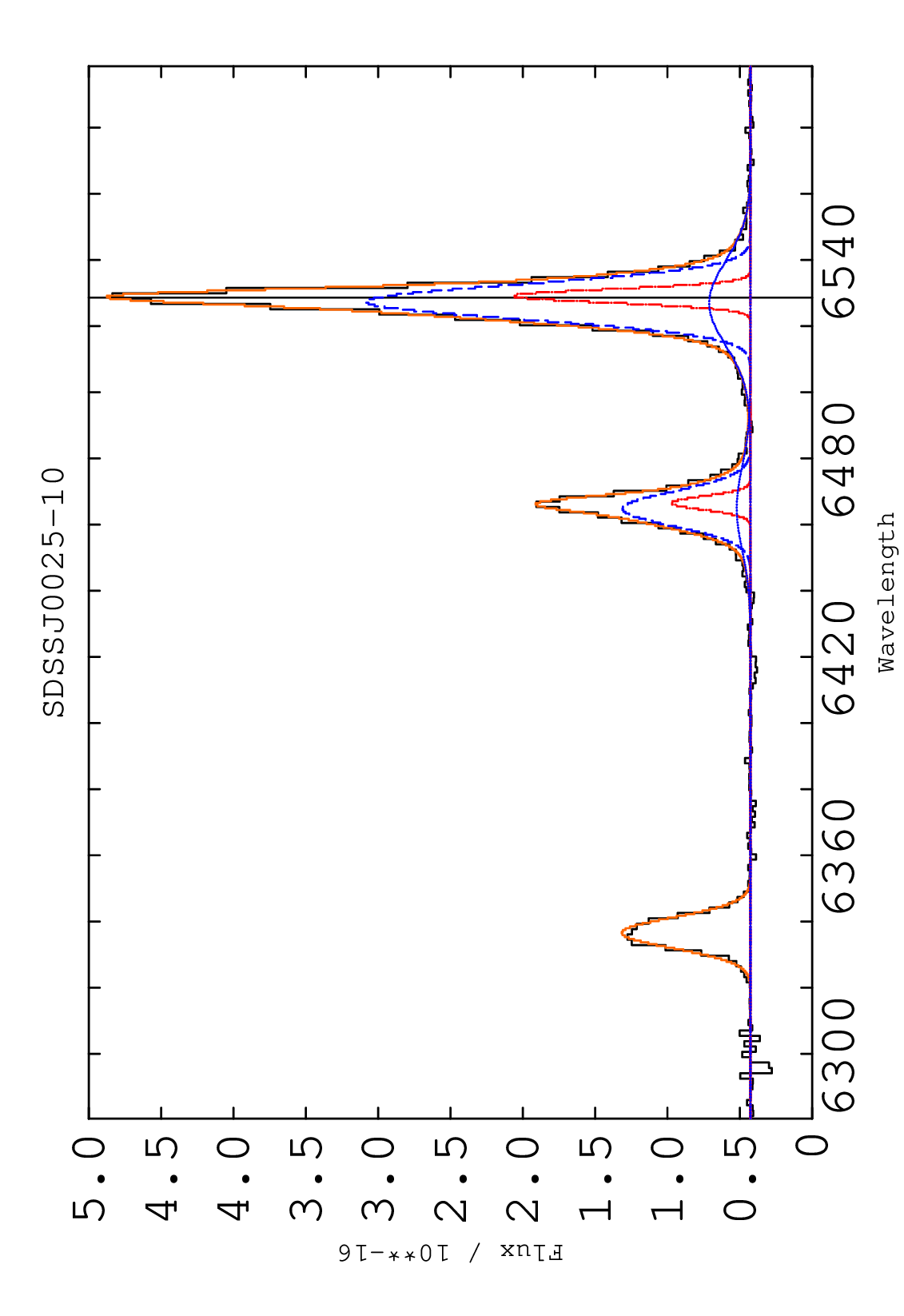}
\vspace{2.8in}
\caption{SDSS J0025-10. Line styles and colors as in Fig.~1.}
\end{figure}

\begin{figure}
\includegraphics{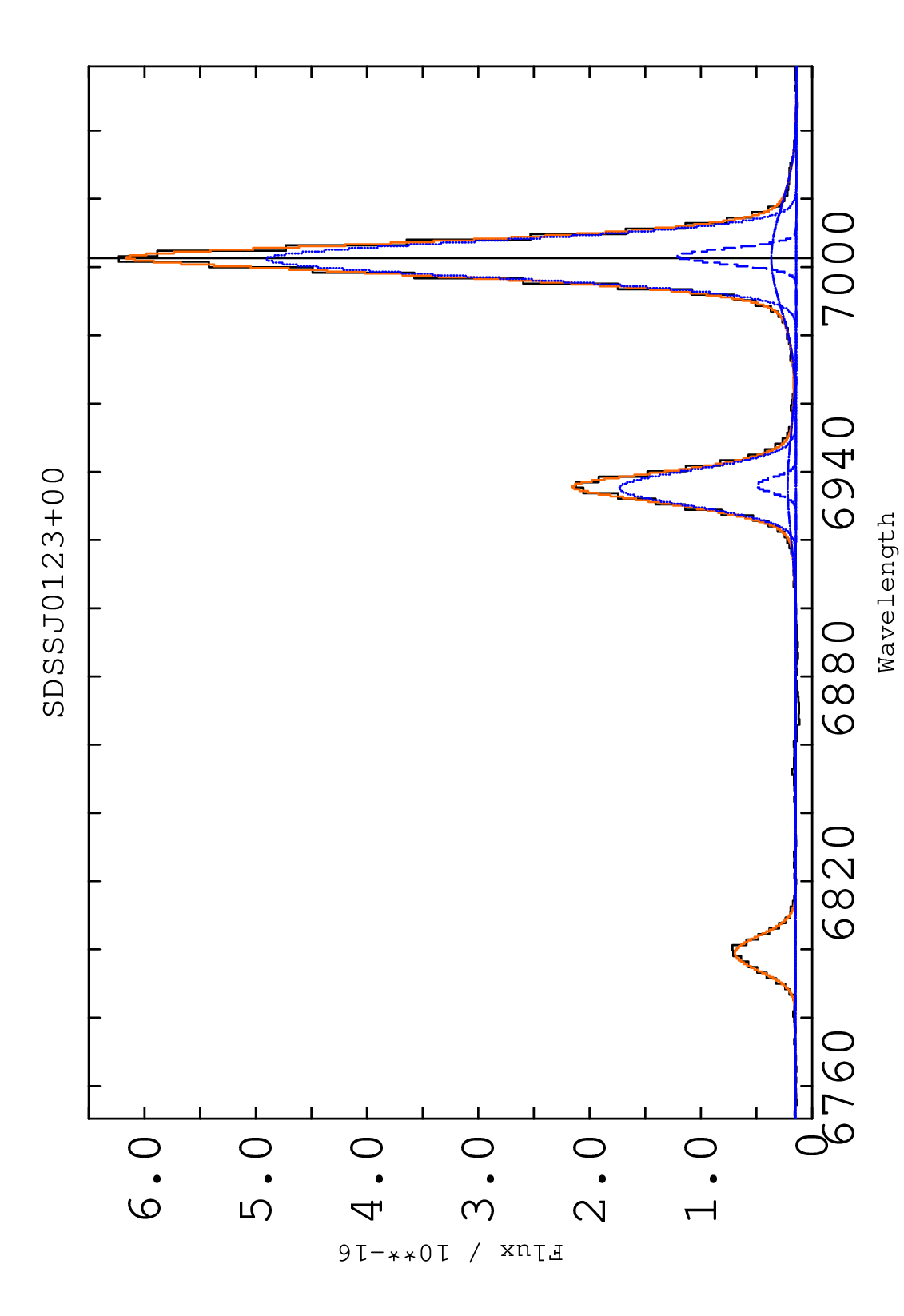}
\vspace{2.8in}
\caption{SDSS J0123+00. Line styles and colors as in Fig.~1.}
\includegraphics{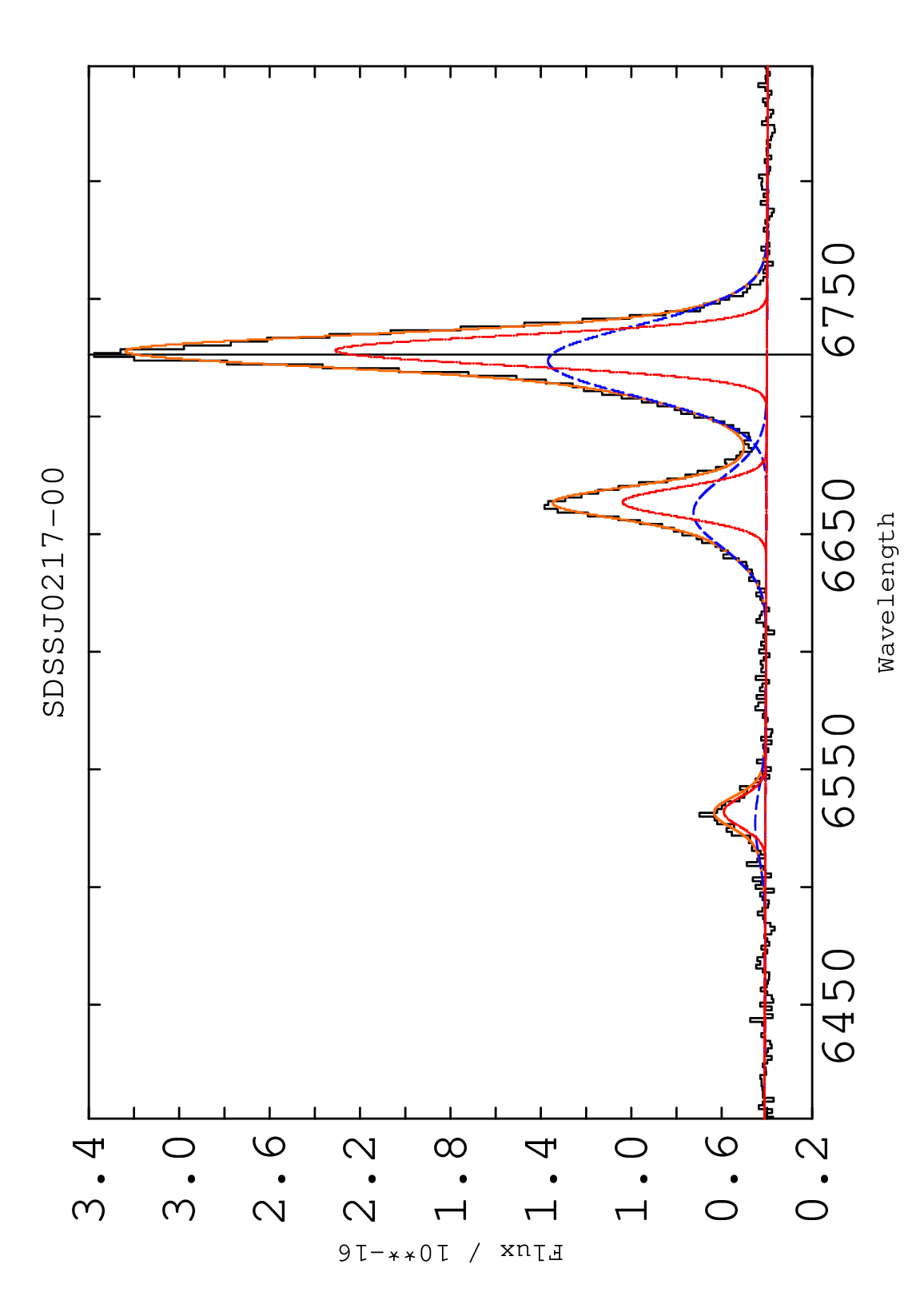}
\vspace{2.8in}
\caption{SDSS J0217-00. Line styles and colors as in Fig.~1.}
\end{figure}

\begin{figure}
\includegraphics{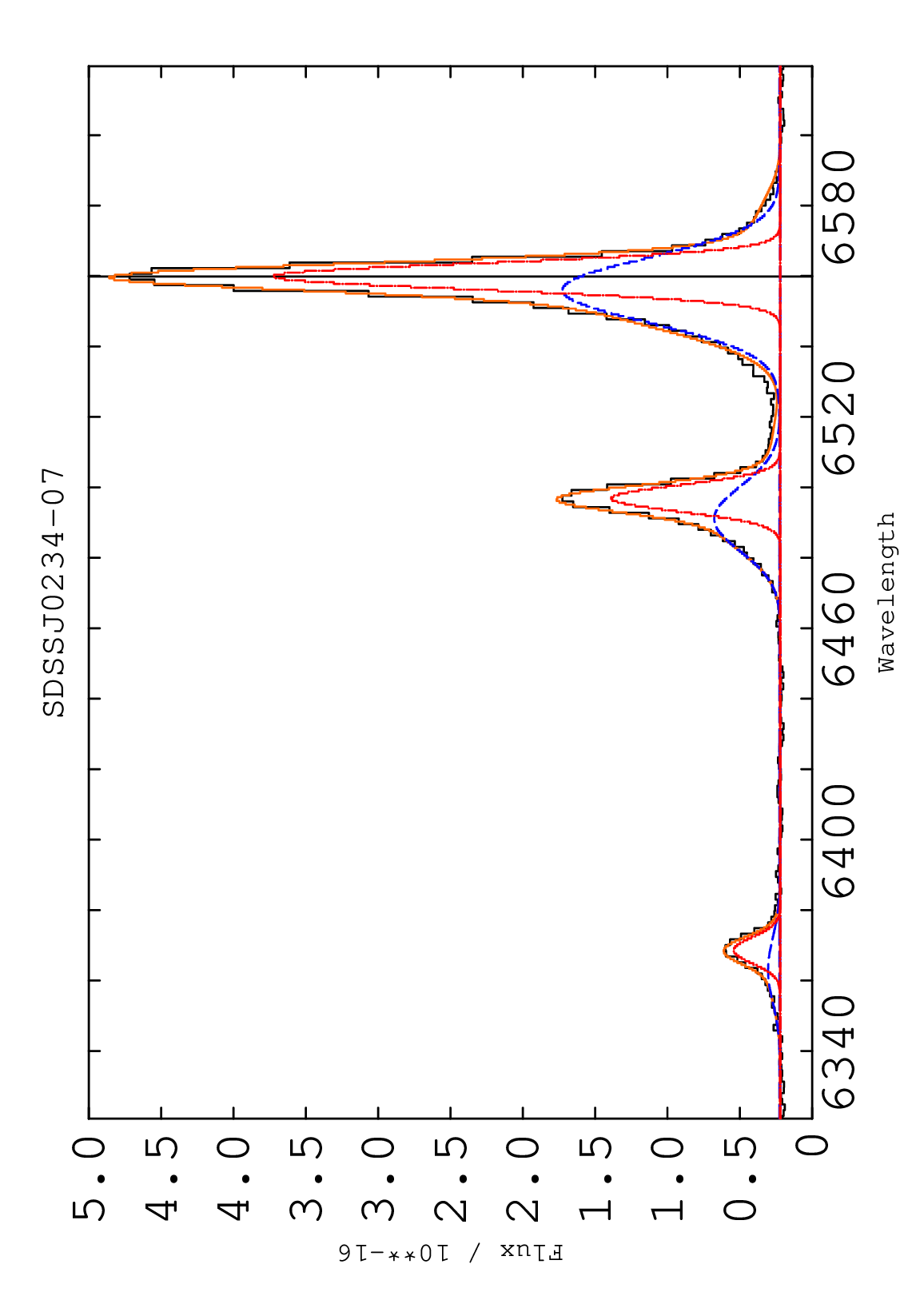}
\vspace{2.8in}
\caption{SDSS J0234-07. Line styles and colors as in Fig.~1.}
\includegraphics{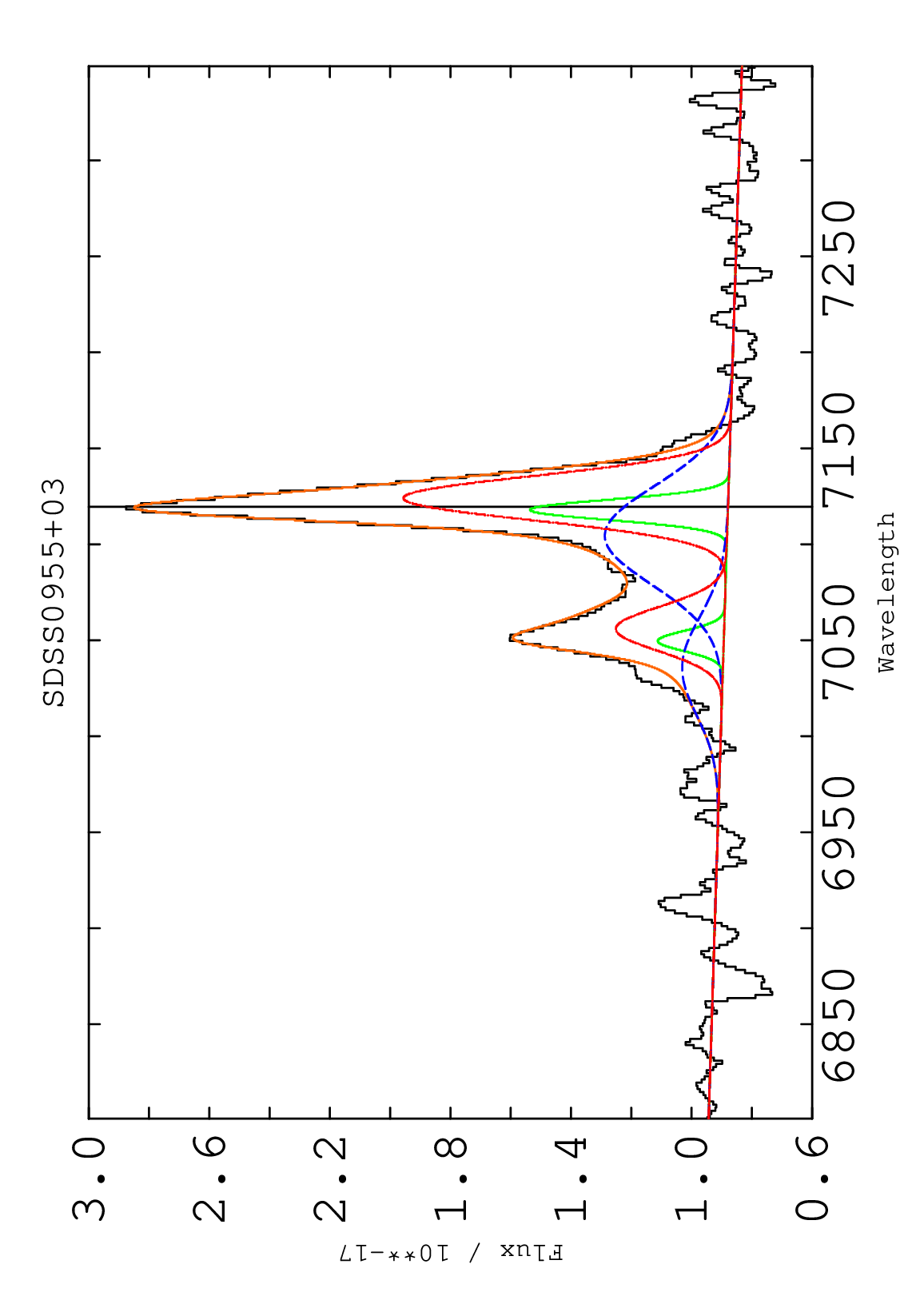}
\vspace{2.8in}
\caption{SDSS J0955+03. Line styles and colors as in Fig.~1.}
\includegraphics{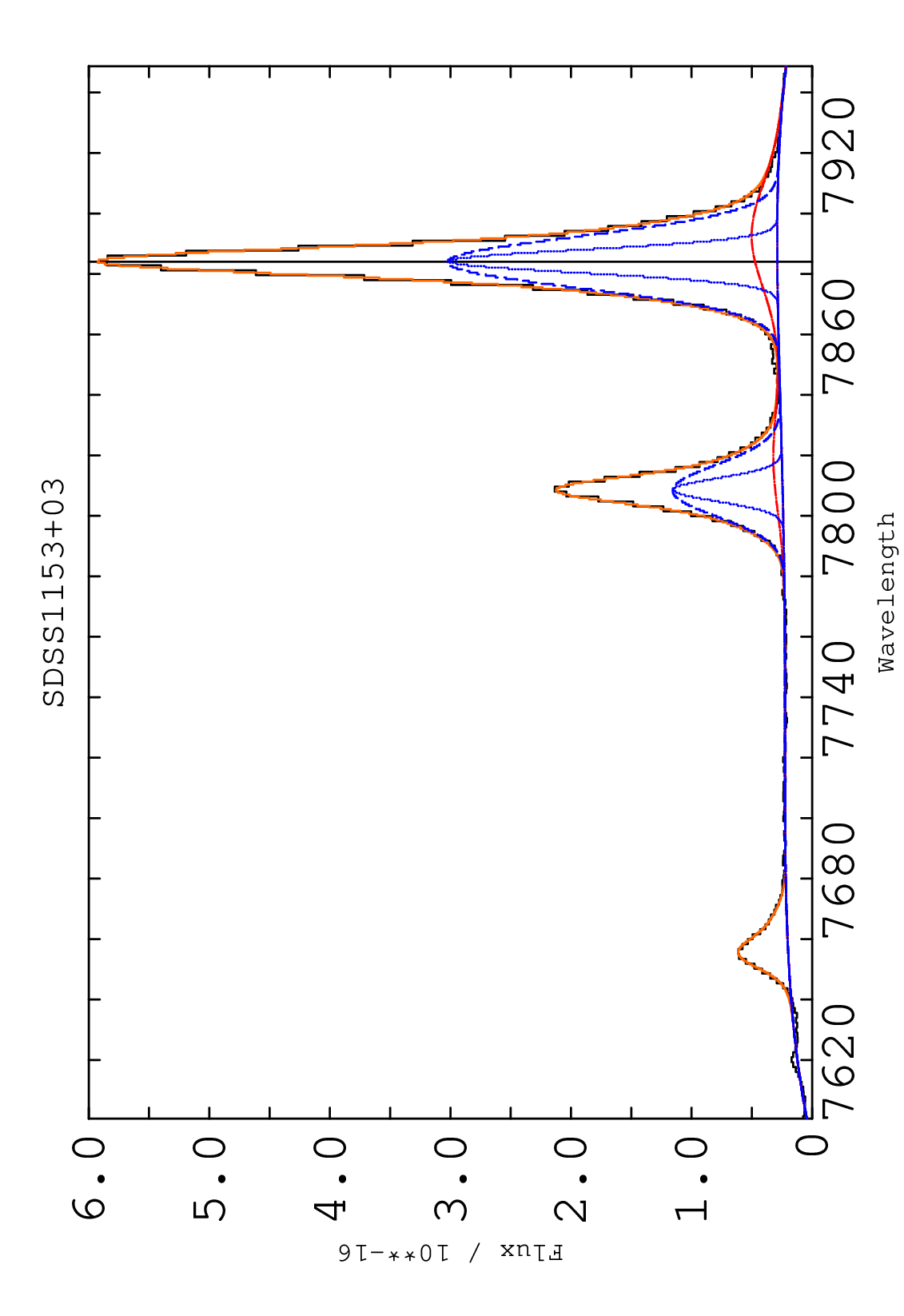}
\vspace{2.8in}
\caption{SDSS J1153+03. Line styles and colors as in Fig.~1.}
\end{figure}

\begin{figure}
\includegraphics{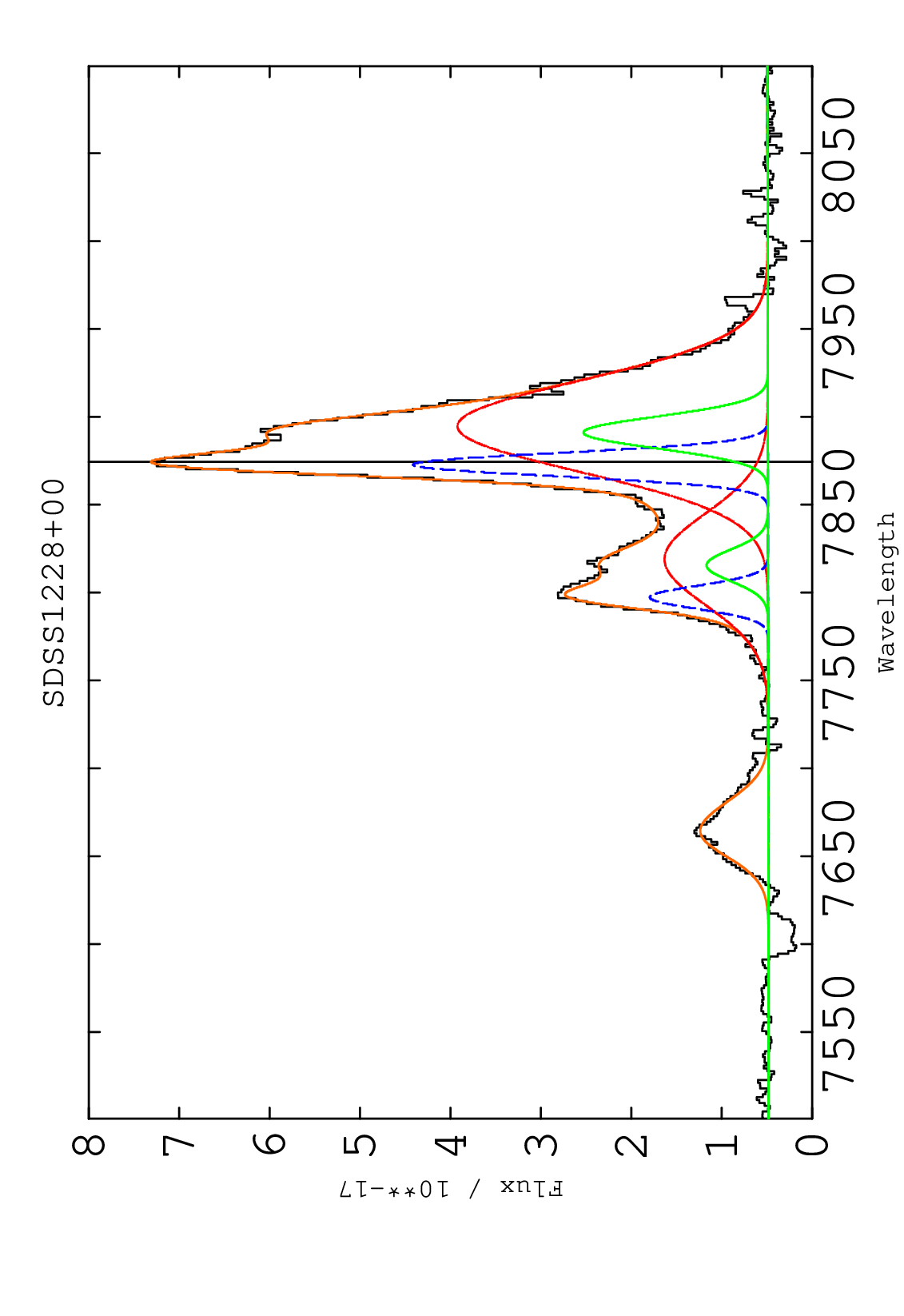}
\vspace{2.8in}
\caption{SDSS J1228+00. Line styles and colors as in Fig.~1.}
\includegraphics{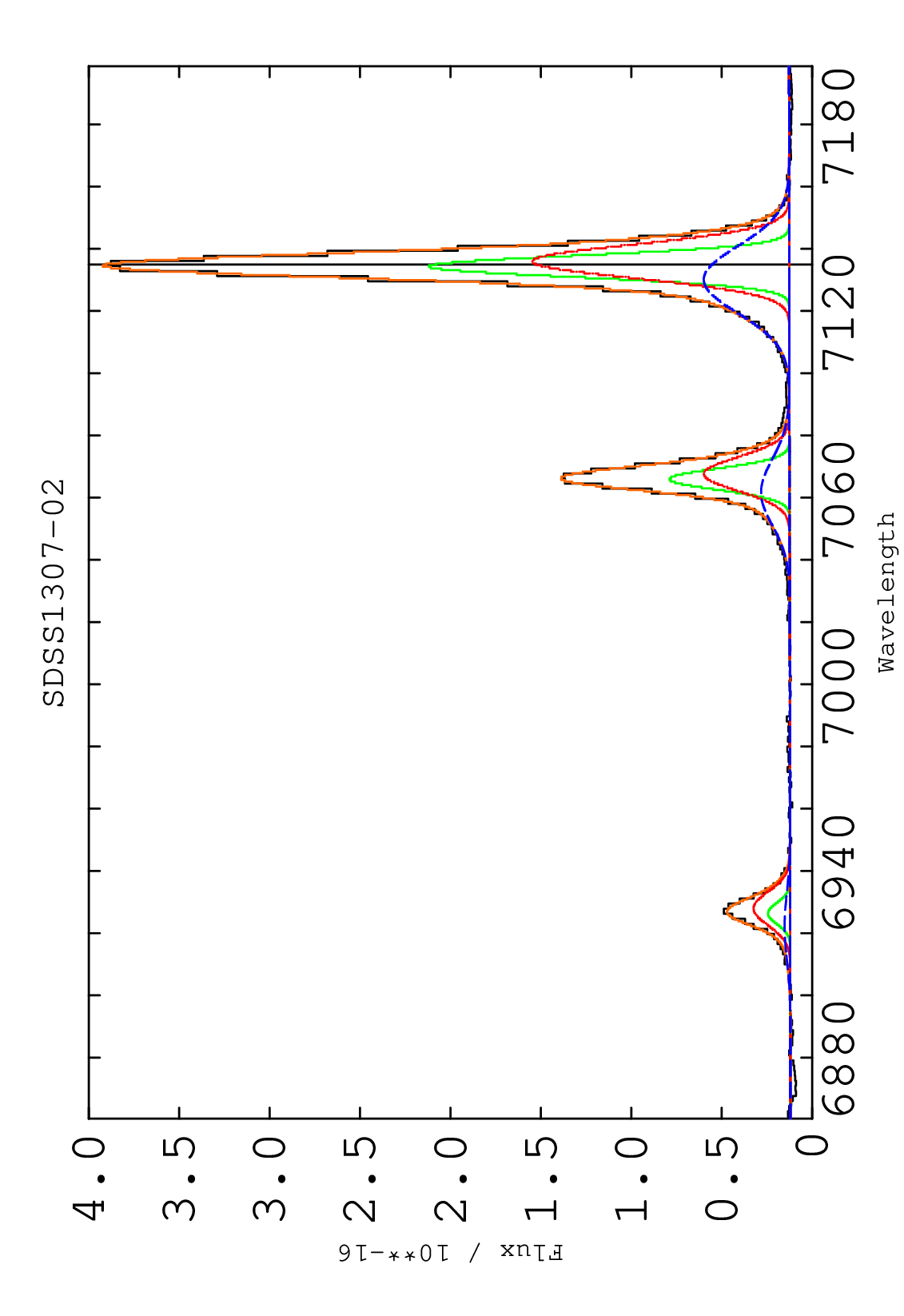}
\vspace{2.8in}
\caption{SDSS J1307-02. Line styles and colors as in Fig.~1.}
\includegraphics{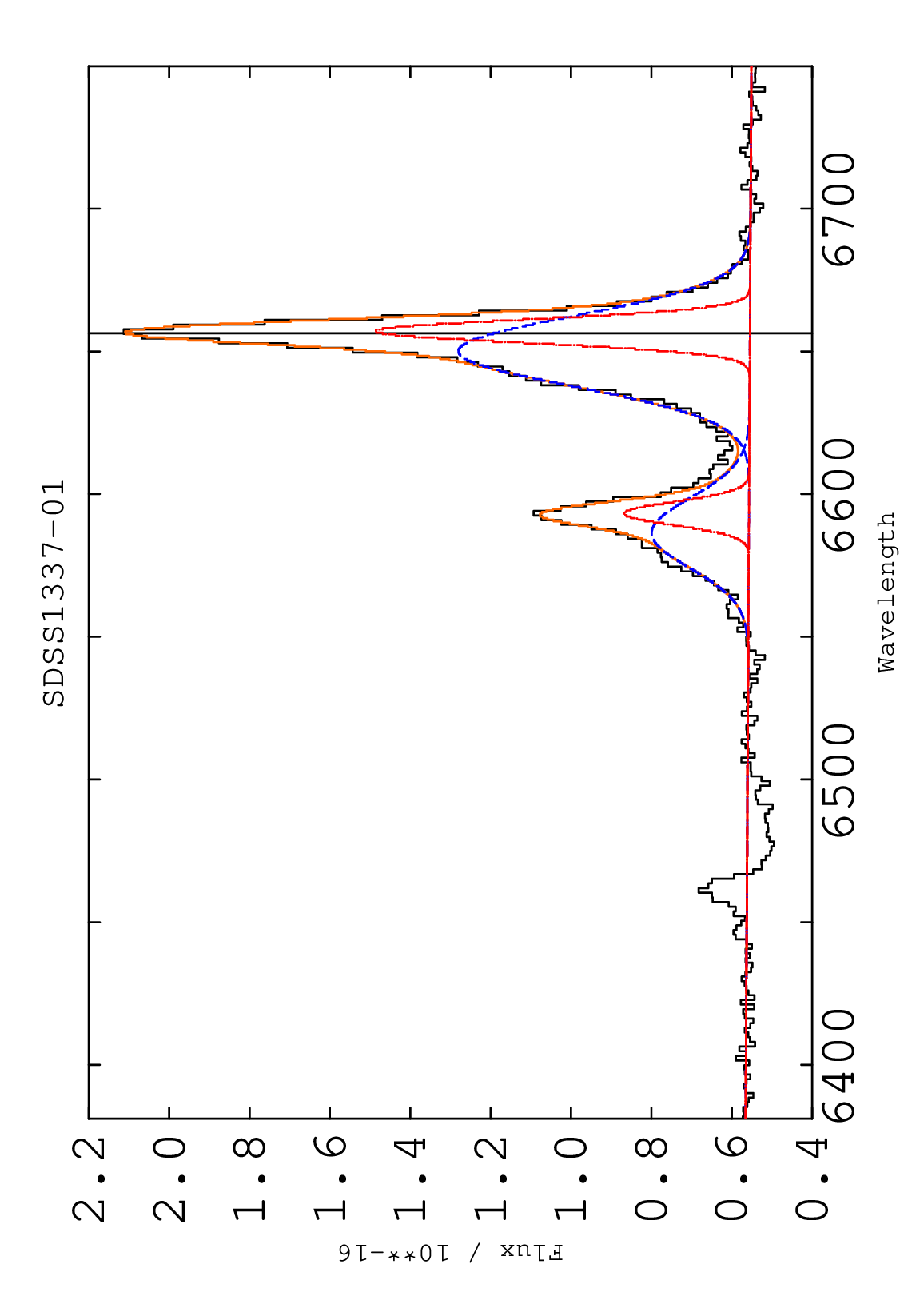}
\vspace{2.8in}
\caption{SDSS J1337-01. Line styles and colors as in Fig.~1.}
\end{figure}

\begin{figure}
\includegraphics{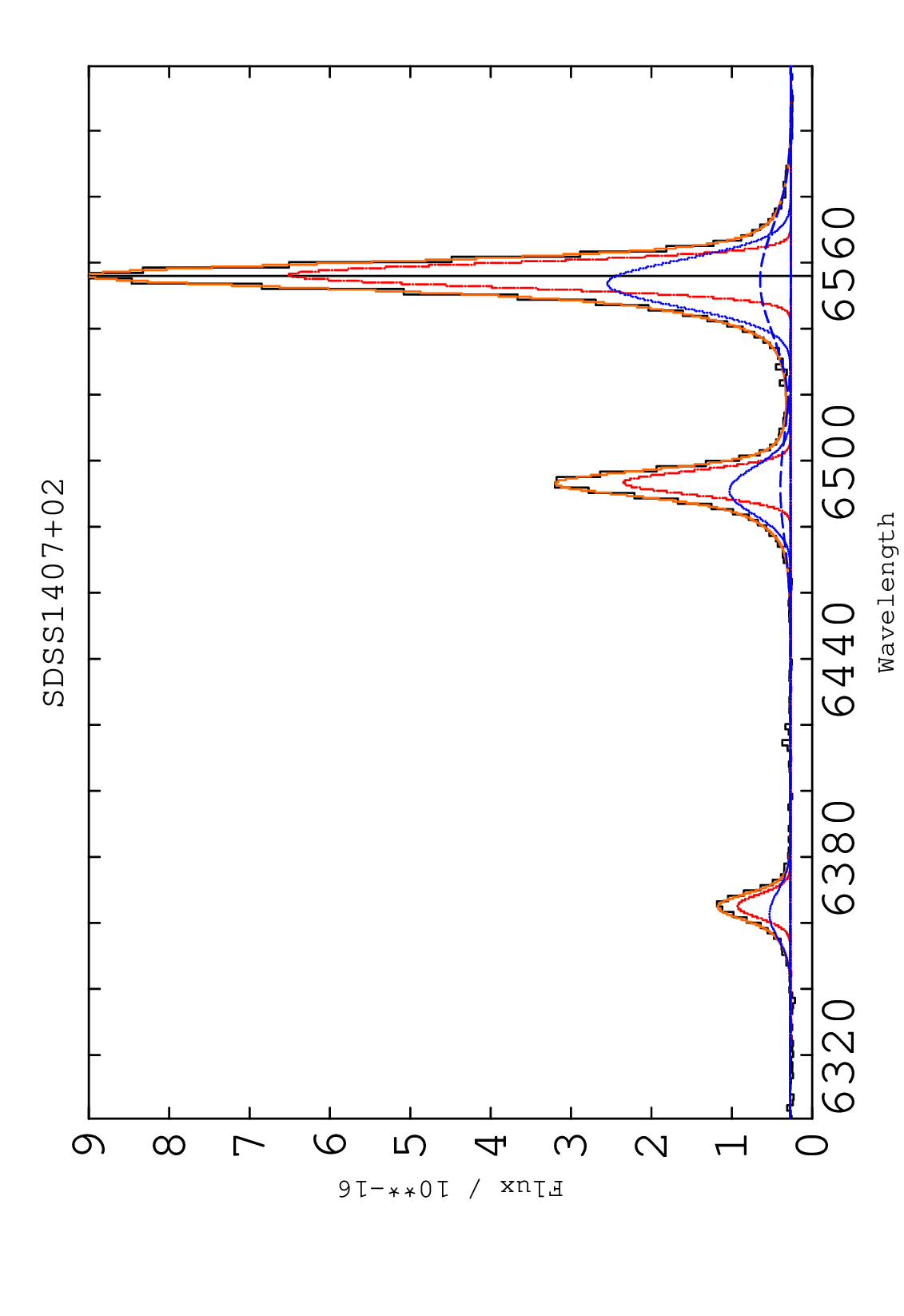}
\vspace{2.8in}
\caption{SDSS J1407+02. Line styles and colors as in Fig.~1.}
\includegraphics{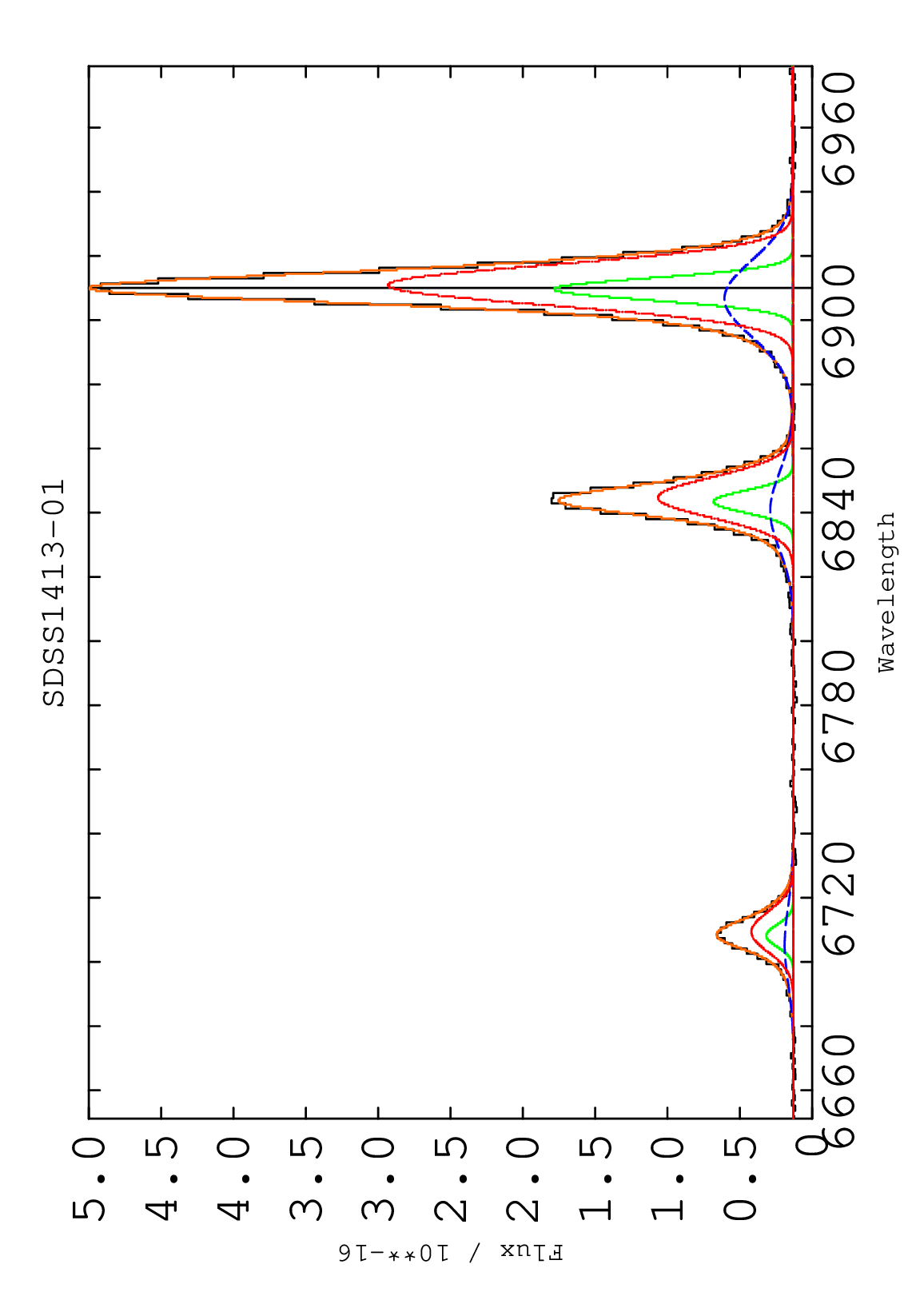}
\vspace{2.8in}
\caption{SDSS J1413-01. Line styles and colors as in Fig.~1.}
\includegraphics{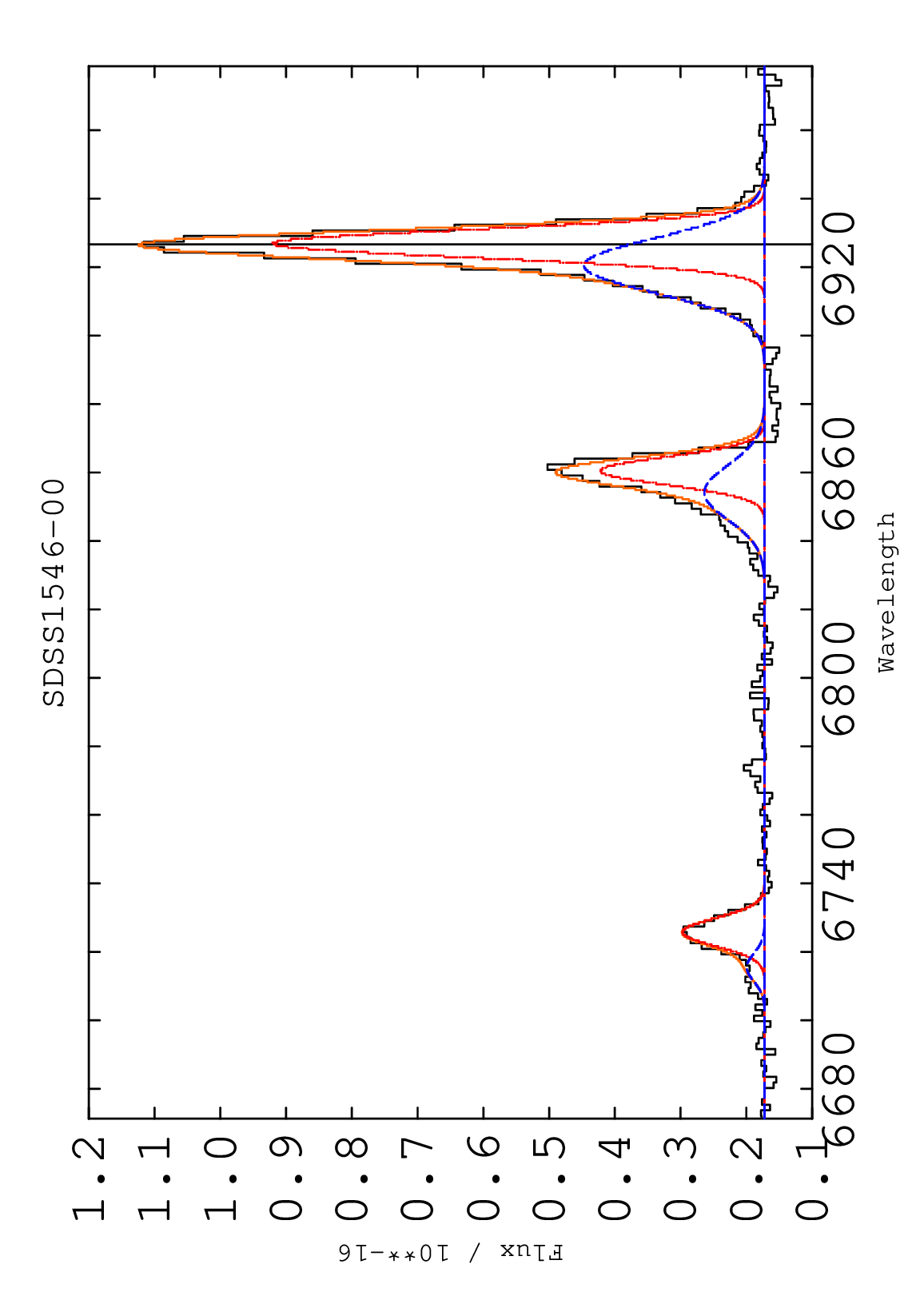}
\vspace{2.8in}
\caption{SDSS J1546-00. Line styles and colors as in Fig.~1.}
\end{figure}

\subsection{The systemic velocity}

In order to characterize accurately the gas motions relative to their environment it is essential to determine the systemic redshift $z_{sys}$ of the galaxy. The safest way to do this is to measure the stellar redshift.  This is often not possible as for example in studies of type 1 AGNs, high redshift  galaxies, etc. where the stellar continuum cannot be isolated from the bright AGN contribution and/or it is not detected with enough signal. In such cases,  alternative methods have been applied. Under the assumption that the BLR gas follows gravitational motions, the broad H$\beta$ line centroid has been used to determine $z_{sys}$. For type 2 objects,
 the [OIII] line fitted with a single Gaussian has often been used   instead,  both to measure $z_{sys}$ and the stellar velocity dispersion.  However, as different works have discussed (e.g. Greene et al. \citeyear{green11}, \citeyear{greene09}, Husemann  \citeyear{huse11}, Boroson \citeyear{bor03}, Boroson  \citeyear{bor05}), this is risky, since the NLR kinematics often reveal non-gravitational motions (see $\delta$1).   Indeed, the very asymmetric and sometimes broad [OIII] spectral profiles  often make the determination of its own $z$  uncertain by up to 200   km s$^{-1}$ (e.g.  Heckman et al. \citeyear{heck81}).

There are three objects in our sample with enough signal in the continuum and detected stellar features to try to estimate $z_ {sys}$: SDSS J0217-00, SDSS J1337-01 (both with strong  stellar features) and SDSS J0123+00 (faint  features).  For this, we use the full spectral fitting method with the Starlight code (Cid Fernandes et al. \citeyear{cid05}), several sets of evolutionary stellar models (Gonz\'alez Delgado et al.  \citeyear{gon05}; Charlot \& Bruzual (in prep.)) and two of the most up to date high/intermediate spectral resolution stellar libraries (Martins et al. \citeyear{mart05}; S\'anchez-Bl\'aquez et at. \citeyear{sanc06}). 

\begin{figure}
\includegraphics{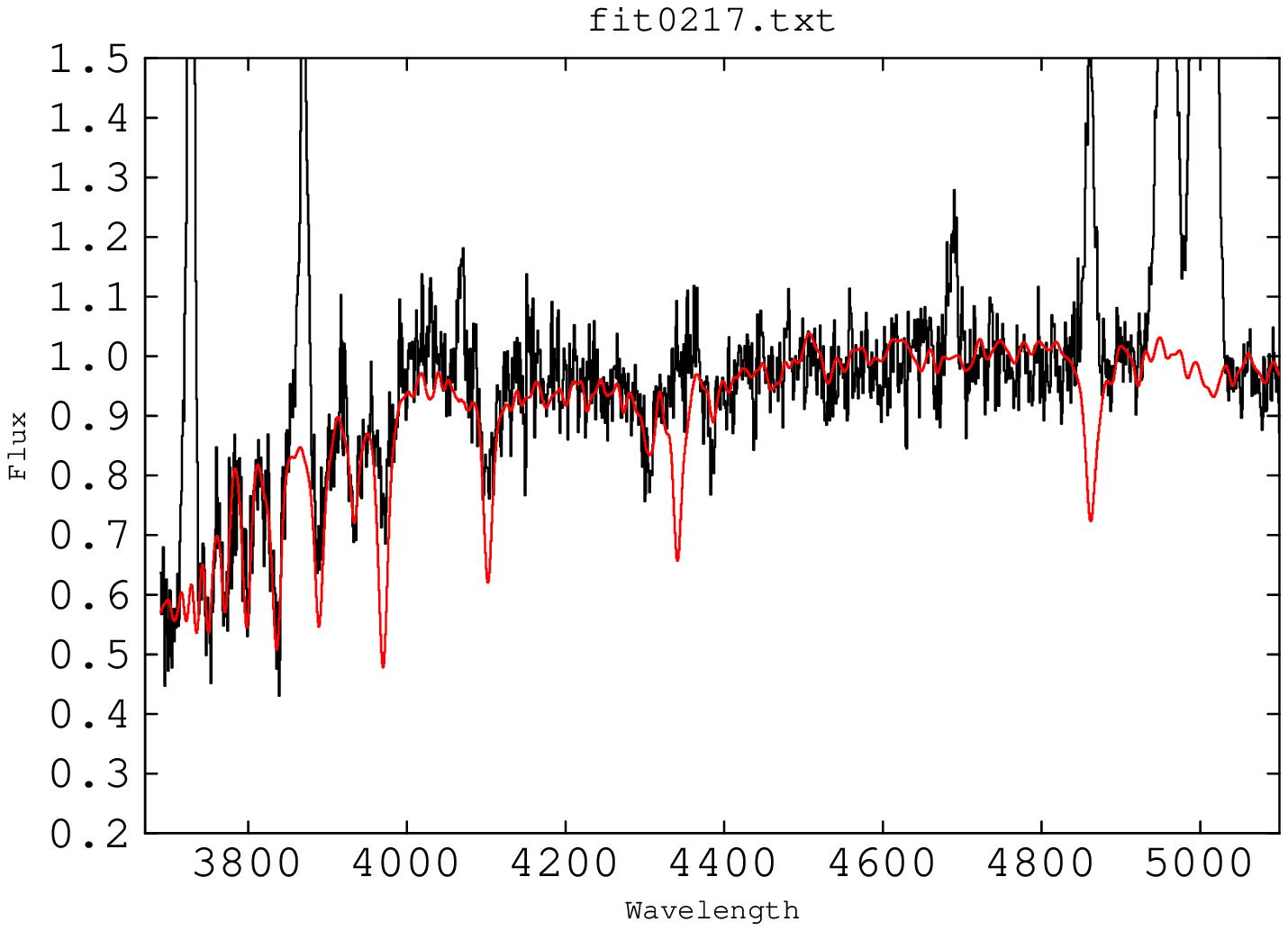}
\vspace{2.8in}
\caption{Fit to the continuum of SDSS J0217-00 using stellar population synthesis models. The spectrum has been translated to  $z$=0. The fluxes have been normalized to the flux at 4020\AA. 
Different templates produce a good fit, so that a range of stellar (and therefore systemic) redshifts is allowed. These  imply a velocity range of $\sim$100 km s$^{-1}$.}
\end{figure}

We find that in all three cases there is a variety of models that produce acceptable fits (see one example in Fig.~14).  These models allow a range of stellar redshifts which depend mainly 
on the single stellar population models used as templates for the fits and the weight given to the G band and CaII K stellar absorption features. The allowed redshifts imply  a velocity range of $\sim$100 km s$^{-1}$, which is the main uncertainty affecting the method.

Although the [OIII] line in the nuclear spectra of our sample is in general quite broad and shows complex kinematic substructure (see $\delta$3), it shows a very narrow core whose redshift can be estimated with an accuracy of several km s$^{-1}$. We will use this as a measurement of $z_ {sys}$.

 In the three objects mentioned above, it is found that this core is at  $\la$100 km s$^{-1}$ of the stellar velocity, the uncertainty being due to the range of stellar redshifts allowed by the  models.  Given the difficulty to infer $z_{sys}$ from the stars in most quasar spectra studied here, we will use  the  narrow [OIII] core in all objects instead. 
This method is supported by the results of \cite{greene05}. They  compared the gaseous and stellar kinematics  in the nuclei of a large sample of active galaxies using spectra of somewhat better resolution (R$=\lambda/\Delta\lambda\sim$1800 vs.  R$\sim$900-1400 at the [OIII] wavelength in our spectra, depending on the object). They  found that  the FWHM of the [OIII] core,  i.e. after the asymmetric wings were removed,  and FWHM$_{stars}$ are correlated (although with a large scatter) according to   FWHM$_{core}$=(1.24$\pm$0.76) $\times$ FWHM$_{stars}$. 

Based on this result, they propose that FWHM$_{core}$  is a reasonable tracer of the stellar velocity dispersion. It seems reasonable to expect that it also provides a good estimation of the stellar redshift and therefore  $z_{sys}$.

\cite{greene09} questioned later  the validity of this result for  luminous  obscured AGNs,
based on the lack of correlation between the [OIII] FWHM and the stellar velocity dispersion. However, the  analysis is based on single Gaussian fits to the emission lines  that ignore their complex kinematic substructure. Given the high complexity of the line profiles (see $\delta$3.2), it is not surprising that no correlation is found. As Table 1 shows, the FWHM of the core is often noticeably smaller than that resulting from a single Gaussian fit to the line profile. 

\subsection{The identification of kinematically perturbed gas}

We are specially interested in those gas components whose kinematic properties suggest non-gravitational, perturbed motions. In order to identify them, we  need to compare  the FWHM of each kinematic component with the stellar velocity dispersion FWHM$_{stars}$. Similar values could be explained if the gas is supported by random motions\footnote{Type 2 quasars are most frequently  associated with elliptical galaxies (Mainieri et al. \citeyear{mai11}, Zakamska et al. \citeyear{zak06}), so random motions are possible}, while rotation  would produce narrower FWHM for the gas than for the stars.  Broader gas components would imply kinematic perturbation. 

In general, it is not possible to measure FWHM$_{stars}$ directly from our spectra. Instead,  following  Greene \& Ho (2005) 
we will infer the value for each object using  FWHM$_{core}$  ($\delta$2.1).  Although the scatter of the correlation between these two parameters produces large uncertainties, the FWHM$_{core}$ values in Table 1 suggest FWHM$_{stars}$ in the range $\sim$250-900 km s$^{-1}$ (Table 2), consistent with values measured in other type 2 quasars (Bian et al. \citeyear{bian07}, Liu et al. \citeyear{liu09}) using stellar features.

The velocity shifts $V_S$ of the different [OIII] kinematic components relative to the narrow line core will  provide additional information to characterize the kinematics.

\section{Results}

\subsection{Integrated line profiles}

We show in Table 1  the FWHM  of [OIII] and H$\beta$ lines. For an initial, general characterization of these quantities we  have applied single Gaussian fits to both lines. The goal of this is to compare with similar studies of type 1 quasars.
 
\begin{table*}
%\tiny
\centering
\begin{tabular}{lllllllll}
\hline
Object  &  z &   FWHM  & FWHM  & FWHM$_{core}$ & AI10 & AI20 & AI50 \\ 
 &   & [OIII] & H$\beta$ &  & & &  \\ \hline
(1)  & (2) & (3) & (4) & (5) & (6) & (7) & (8) \\ \hline
SDSS J2358-00  &  0.402 &   440$\pm$50  & 420$\pm$85  & 430$\pm$10 & 0.03$\pm$0.04 & 0.00$\pm$0.03 & 0.00$\pm$0.03  \\
SDSS J0025-10  & 0.303 &   435$\pm$55& 420$\pm$70 & 320$\pm$80 & {\bf 0.33$\pm$0.05} & {\bf 0.36$\pm$0.08}   & {\bf 0.42$\pm$0.09}\\
SDSS J0123+00  & 0.399  &  445$\pm$45 &  470$\pm$60  & 430$\pm$50 & {\bf 0.10$\pm$0.05} & 0.07$\pm$0.07 &  0.04$\pm$0.06  \\
SDSS J0217-00 & 0.344  &  910$\pm$30 &  700$\pm$60  & 700$\pm$60 & {\bf 0.22$\pm$0.03} & {\bf 0.06$\pm$0.04} & {\it -0.17$\pm$0.09}    \\
SDSS J0234-07 & 0.310  &   510$\pm$60 & 420$\pm$60 & 450$\pm$50 &  {\bf 0.39$\pm$0.04} & {\bf 0.52$\pm$0.05} &   {\bf 0.53$\pm$0.09} \\
SDSS J0955+03  &  0.422  &    1320$\pm$30 & N/A  & 940$\pm$60  &  {\bf 0.13$\pm$0.03} & {\bf 0.14$\pm$0.07} &  {\it -0.15$\pm$0.09} \\
SDSS J1153+03 & 0.575  &  550$\pm$10  &   N/A   &440$\pm$20 & -0.05$\pm$0.06 & -0.05$\pm$0.04 &   -0.06$\pm$0.09 \\
SDSS J1228+00 & 0.575 &   1860$\pm$20   & N/A & 1100$\pm$100 & {\it -0.66$\pm$0.03} & {\it -0.41$\pm$0.04} & {\it -0.49$\pm$0.05}   \\
SDSS J1307-02 & 0.455 &      410$\pm$10  &  455$\pm$15 & 360$\pm$10  & {\bf 0.15$\pm$0.03} & {\it -0.10$\pm$0.05}  &   0.00$\pm$0.04 \\
SDSS J1337-01 &  0.329   &    900$\pm$30  & N/A & 540$\pm$50 & {\bf 0.33$\pm$0.07}  & {\bf 0.19$\pm$0.07}  & {\bf 0.15$\pm$0.06}    \\
SDSS J1407+02 &  0.309 &    430$\pm$20 & 460$\pm$15 & 400$\pm$10 & {\bf 0.18$\pm$0.06}    & {\bf 0.17$\pm$0.09} & {\bf 0.20$\pm$0.10} \\
SDSS J1413-01   &  0.380 &    550$\pm$10 &  570$\pm$40 & 500$\pm$10    &  0.03$\pm$0.04 & 0.03$\pm$0.03 & 0.00$\pm$0.03  \\
SDSS J1546-01 & 0.383 &    500$\pm$40 &  440$\pm$30 & 430$\pm$20  &  {\bf 0.39$\pm$0.08} &  {\bf 0.43$\pm$0.08} & 0.00$\pm$0.05 \\  \hline
Median & 0.383  &    510   &  455 &   440  &  {\bf 0.13} & {\bf 0.07} & 0.00 \\ 
STD  &   0.091 &    440  &  93 & 230   & 0.28 & 0.24 & 0.26 \\  \hline
\end{tabular}
\caption{Type 2 quasar sample studied in this work with kinematic measurements based on single Gaussian  fits
to the [OIII] and H$\beta$ lines.  The redshift is shown in column (2). The  [OIII] and H$\beta$ FWHM are are in columns (3) and (4). Column (5) gives the FWHM of the narrow [OIII] core that remains after the asymmetric line wings are removed.   AI10, AI20 and AI50 in columns (6), (7) and (8) quantify the degree of asymmetry of the [OIII] line profile at 10\%, 20\% and 50\% intensity levels respectively. Positive, negative and cero values (taking errors into account) correspond to blue, red and no asymmetries, which are highlighted in bold, italic and normal fonts respectively.} 
\end{table*}

  The [OIII] FWHM values in Heckman, Miley \& Green (\citeyear{heck84}) subsample of 35 type 1 quasars are in the range [260-850] km s$^{-1}$, with a median value of 460 km s$^{-1}$. 
 If only the 10 radio quiet quasars are considered the range is the same, but the median value increases to 570 km s$^{-1}$.
The range in our sample is [410$\pm$10 to 1860$\pm$20] km s$^{-1}$ and the median is   510 km s$^{-1}$.  
 
A visual inspection of the [OIII] spectral profiles reveals that it is asymmetric in the majority of objects.
 Following \cite{heck81}, we have quantified the degree of asymmetry at 10\%, 20\% and 50\% of the line maximum intensity using the parameters AI10, AI20 and AI50. AI10 is defined as $\frac{WL10-WR10}{WL10+WR10}$, where WL10 and WR10 are the half  widths to the left and right  of the center of the narrow [OIII] core  at 10\% intensity level. The widths were corrected for instrumental broadening. AI20 and AI50 are defined in the same way, but at 20\% and 50\% intensity levels respectively. AI10 is introduced compared with previous studies (see also Spoon \& Holt \citeyear{spoo10}) because we find that in most objects the asymmetry is perceived at very low intensity levels, in the form of faint wings. The schematic representation of the various parameters involved in these calculations is shown in Fig.~15. The half widths WL10', WR10', etc have a `` ' " because they correspond to values not corrected for instrumental broadening.

AI10, AI20 and AI50 are shown in Table 1. Positive, negative and 0 values indicate blue (highlighted in bold), red  (italic fonts) and no asymmetries (normal fonts)  respectively. It can be seen that the fraction of asymmetric profiles increases at lower intensity levels (7/13 at 50\%; 9/13 at 20\% and 10/13 at 10\% levels). While at 50\% level there is a variety of asymmetries with no clear preference for a given sign, at lower intensity  blue asymmetries become predominant: at 10\% level 9/13 objects show a blue asymmetry, vs. 2 with red and 2 more which no asymmetry. 

   Our results are consistent with \cite{heck84}, who found a clear predominance of blue asymmetric [OIII] profiles in radio quiet and flat steep spectrum (compact) radio loud quasars  at $z\le$0.5. The asymmetries disappear in steep spectrum (extended) radio loud objects. None of the objects in our sample are radio loud. 
       
 \begin{figure}
\includegraphics{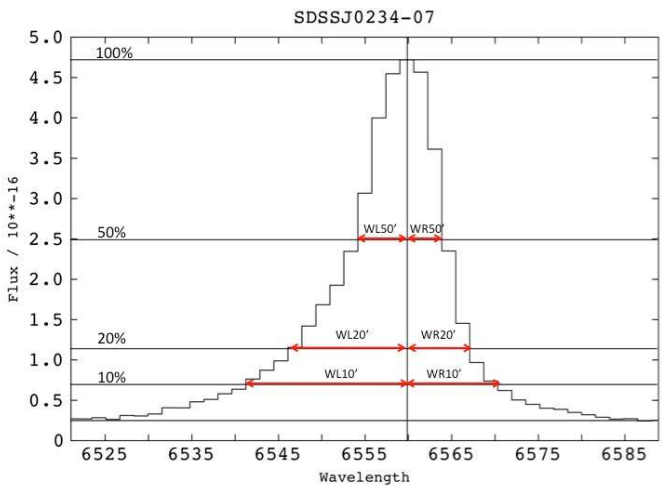}
\vspace{2.7in}
\caption{Schematic representations of the various parameters used to quantify the degree of asymetry of the [OIII] line profile at differente intensity levels, following Heckman et al.  (1981).}
\end{figure}

 An excess of blue asymmetries suggests that a significant fraction of the line emitting gas  is moving radially  in an outflow (see $\delta$1).

\subsection{Spectral decomposition of the H$\beta$-[OIII] line profiles.}

The [OIII] spectral line profiles show complex kinematic substructure   in all 13 objects:  3 kinematic components are required to produce the best fit in 8/13 objects, while 2 are required for the remaining 5 (Fig. 1 to 13). 

All quasars in our sample have one or two kinematic components with [OIII]  FWHM $\la$FWHM$_{stars}$ (Table 2), i.e., they show no sign of kinematic disturbance. The presence of  two components which seem to follow gravitational motions is not surprising, since the emissions from the NLR (probably dominant) and possibly also the EELR are expected within the spatial apertures used
to extract the spectra.  According to studies of narrow line radio galaxies,  the EELR is expected to have different kinematic, physical and ionization properties than the NLR (e.g. Tadhunter, Fosbury \& Quinn 
\citeyear{tad89}, Robinson et al. \citeyear{rob87}).

{\it All} objects show, moreover, a very broad component with [OIII] FWHM$\ga$2$\times$FWHM$_{stars}$.  In general,  this component is  broader  than the maximum possible value of   FWHM$_ {stars}$  allowed by \cite{greene05} correlation scatter (see $\delta$2.2.). Therefore, the uncertainty on FWHM$_{stars}$ does not affect
our conclusion.  Thus, the NLR of all type 2 quasars in our sample contains highly perturbed gas, in addition to the more quiescent, ambient gas whose FWHM is consistent with gravitational motions.

The FWHM of the perturbed gas components are in the range 780$\pm$30  to 2500$\pm$200 km s$^{-1}$ with a median value of 1280 km s$^{-1}$. 
Their velocity shifts  relative to the  [OIII] narrow core $V_S$  are in the range  10$\pm$40  and  up to $\sim$-620$\pm$140 km s$^{-1}$, with a median value of $\sim$-145 km s$^{-1}$ (see also Greene et al. \citeyear{green11}). 
 
\begin{figure}
\includegraphics{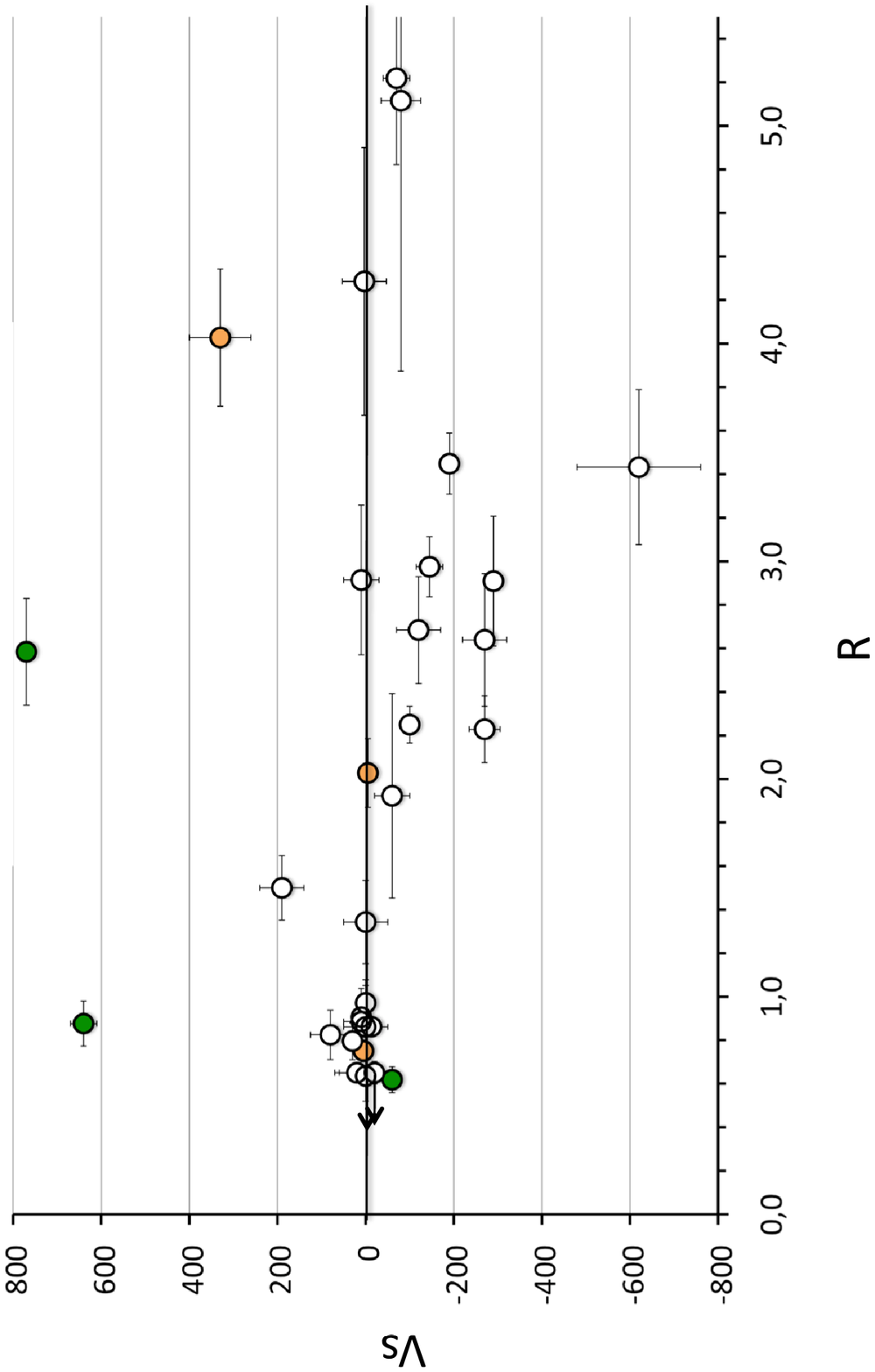}
\vspace{2.8in}
\caption{Velocity shift   $Vs$ in km s$^{-1}$ vs. $R=\frac{FWHM[OIII]}{FWHM _{stars}}$ for all kinematic components isolated in the spectral fits. Coloured symbols correspond to the two peculiar objects suspected to have intermediate orientation between type 1 and type 2 (green for SDSS J1228+00; orange for SDSS J1153+03). Excluding these, there is a global trend for the perturbed gas components ($R\ga$2) to be bluesfhited, while more quiescent kinematic components $R\la$1) show small or null velocity shifts.}
\end{figure}

We have plotted $R=\frac{FWHM[OIII]}{FWHM_{stars}}$ vs. $V_S$ for all kinematic components (Fig. 16).
The two  objects highlighted in colour are  SDSS J1153+03 (orange) and SDSS J1228+00 (green). \cite{vil11} proposed that these quasars might have an intermediate  orientation  between type 1 and type 2, so  that we have a direct view of an intermediate density region, more interior and closer to the black hole than the NLR we see in the rest of the sample. 
Fig.~16 shows  that they are also peculiar from the kinematic point of view, since they are the only ones showing kinematic components with large red velocity shifts. 

If these two peculiar objects are excluded, Fig.~16 shows a strong trend for the highly perturbed gas (i.e. kinematic components
with $R\ga$2) to be {\it blueshifted}, while the more quiescent gas (components with $R\la$1.0) shows in general no or small velocity shifts. 

 $V_S$ is plotted vs. the FWFHM in Fig.~17 for the individual kinematic components isolated in each quasar (the two peculiar quasars are excluded). The same symbol is used for each object. 
Selecting a given symbol, it can be seen that the largest
FWHM values (i.e. the most perturbed gas in a given object) are in general   associated with the largest blueshifts  (see also Table 2). This result is found in all objects  where the velocity uncertainties allowed a definitive conclusion (7/11).  
 
Thus, the perturbed gas
shows both  broad FWHM compared with the stellar velocity dispersion and a trend to show the largest blueshift. The perturbed gas is therefore responsible for the blue line profile asymmetries.  All together suggest  the presence of ionized gas outflows in the inner $\la$few kpc, such that the receiding gas is obscured by dust.

\begin{figure}
\includegraphics{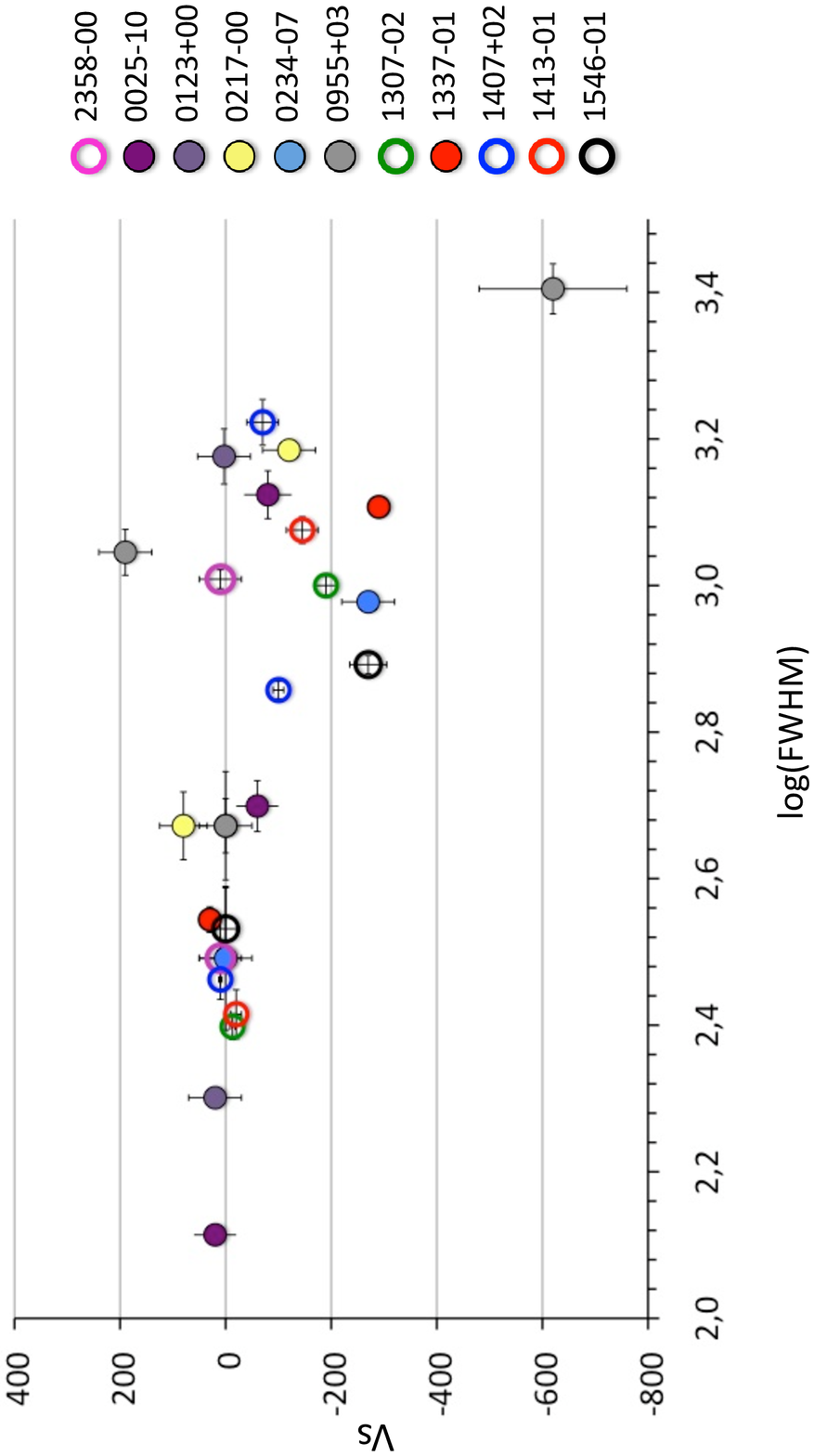}
\vspace{2.55in}
\caption{Velocity shift  $Vs$ (km s$^{-1}$) vs. the FWHM (km s$^{-1}$) in log of all kinematic components isolated in the the [OIII] line for all quasars in the sample, except the two peculiar objects
SDSS J1228+00 and SDSS J1153+03.   Equal symbols correspond to the same object.  There is a general  trend such that in each object the broadest component shows also the largest blueshift.}
\end{figure}

\begin{figure}
\includegraphics{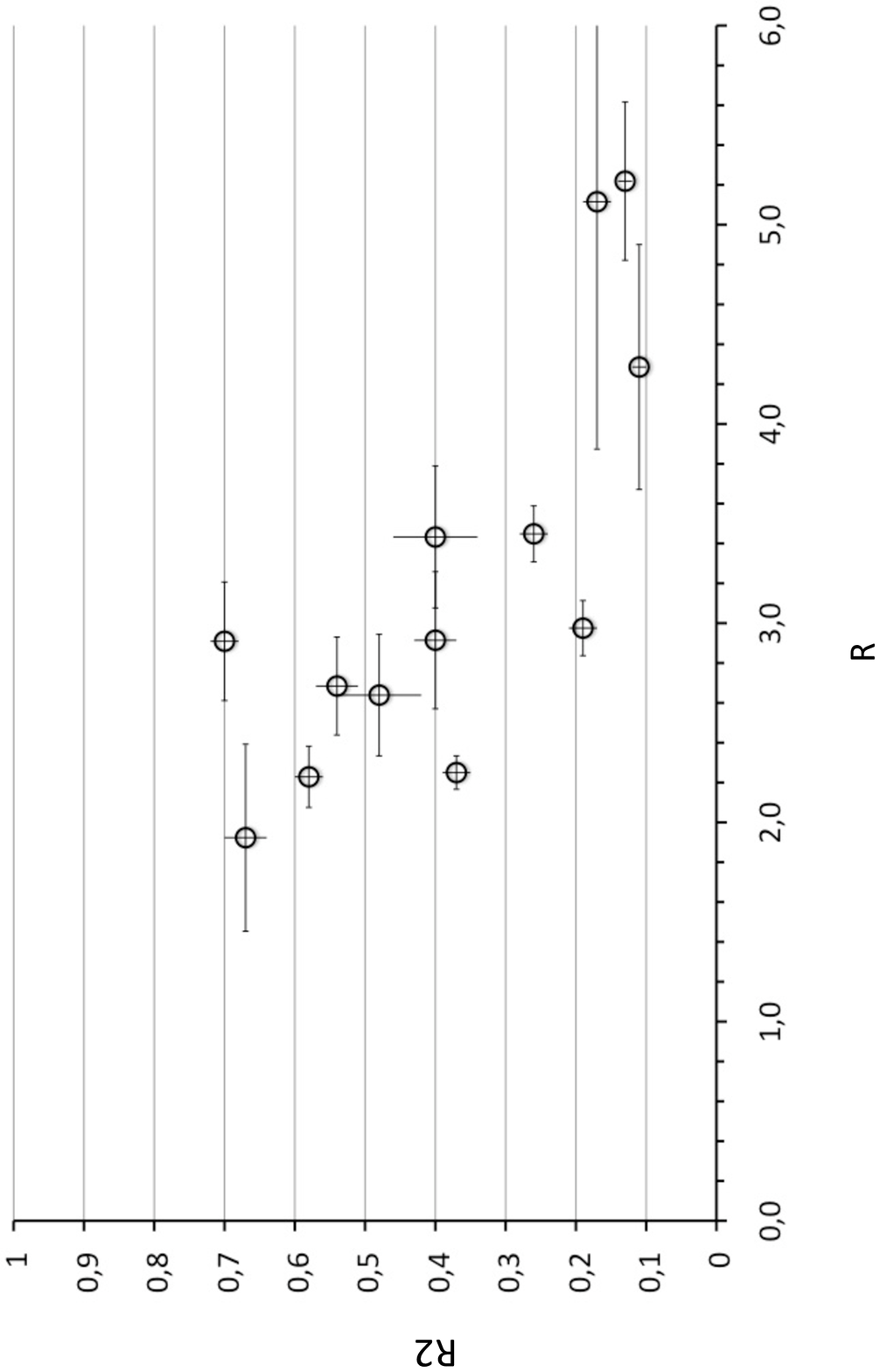}
\vspace{2.9in}
\caption{Relative contribution $R2$  to the total [OIII] line flux of the perturbed component (column 5 in Table 2) vs. $R=$FWHM/FWHM$_{stars}$ for the 
perturbed kinematics components (i.e. the outflowing gas). An anticorrelation is observed. Kinematic components with more extreme kinematics (larger $R$)  have a smaller relative contribution 
to the total line flux.}
\end{figure}

The relative contribution of the perturbed (i.e. outflowing) gas to the total [OIII] flux $R2=\frac{[OIII]^{outf}}{[OIII]^{tot}}$ (column  5 in Table 2) varies strongly from object to object. It can account for $\sim$10\% and up to $\sim$70\% of the total line flux.  
$R2$  is plotted vs. $R$ in Fig. 18. An anticorrelation is found such that, the more perturbed the outflowing gas is (larger $R$), the lower its relative contribution  to the total [OIII] flux. A Spearman's rank correlation coefficient $\rho$=-0.79
is found and a statistical significance at the 99.95\% level.

The perturbed gas is also detected in    H$\beta$ in  7 objects. The  [OIII]/H$\beta$ ratio is in general very high  ($\ge$10, in most cases where it could be measured), which is similar or even larger than that of the narrower kinematic components.  This suggests very high excitation level of the outflowing gas.

Given the size of the apertures used to extract the spectra, the effects of the outflows we see are constrained within  radial spatial scales of $r\la$few kpc from the central engine.  Larger spatial scales of the total outflowing region are not discarded   (e.g. Humphrey et al. \citeyear{hum10}).

\begin{table*}
\tiny
\centering
\begin{tabular}{lrrrrrrrrrrrrr}
\hline
Object &  Asymm & Flux[OIII]   &  $\frac{[OIII]_i}{[OIII]_{tot}}$  & $\frac{H\beta_i}{H\beta_{tot}}$  & $\frac{[OIII]_i}{H\beta_i}$ & FWHM[OIII]  & $V_S$  & $\frac{[OIII]_{tot}}{H\beta_{tot}}$  & L[OIII]$_ {tot}$   & FWHM$_{stars}$ \\
 &  & $\times$10$^{-16}$ & & &  &  &   &   &  $\times$10$^{42}$ \\ 
 &     & erg s$^{-1}$ cm$^{-2}$ & & &  & km s$^{-1}$ &   km s$^{-1}$   &   &   erg s$^{-1}$ & km s$^{-1}$ \\ 
(1) & (2)  & (3) & (4) &  (5) & (6) & (7) & (8) & (9) & (10)  & (11)   \\ \hline
2358-00  & No &  75$\pm$6 &  0.60$\pm$0.06 & 0.62$\pm$0.08     &  10$\pm$1.5  &  310$\pm$40 & 10$\pm$40 & 10.9$\pm$0.4  & 8.06  & 350$\pm$40 \\
 & &   50$\pm$2  &   0.40$\pm$0.03 &  0.38$\pm$0.05  & 11$\pm$1 &    1020$\pm$30  &    10$\pm$40 &  &  \\ \hline
0025-10 & Blue &    8.3$\pm$0.4 & 0.16$\pm$0.01   &   $\le$0.08 &   $\ge$9  & $\le$130 & 20$\pm$40  &   4.6$\pm$0.2  & 2.07 & 260$\pm$60  \\ 
 &   &   35$\pm$1  &  0.67$\pm$0.03     &  $\sim$1    &  3.4$\pm$0.1  & 500$\pm$40 & -60$\pm$40 \\ 
  &  &  8.8$\pm$0.9  &   0.17$\pm$0.02   &    $<$0.17    & $>$4.3 & 1330$\pm$100 &   -80$\pm$45 &  &\\ \hline
0123+00  & Blue &    5.9$\pm$0.9 &  0.08$\pm$0.01  &  $<$0.08    &  $>$8.3      & $\leq$200 &   20$\pm$50 & 10.3$\pm$0.3  & 5.21 & 350$\pm$40   \\ 
 &   &  61.7$\pm$0.6  &   0.81$\pm$0.02   & $\sim$1       &  8.8$\pm$0.2 & 470$\pm$40 &  -15$\pm$50 \\ 
  &  &  8.2$\pm$0.7  &  0.11$\pm$0.01 & $\leq $0.13& $>$7 &1500$\pm$130    &  3$\pm$50       \\ \hline
0217-00  & Blue &    30$\pm$2 &  0.46$\pm$0.03   & 0.63$\pm$0.09    & 10$\pm$2    & 470$\pm$50  & 80$\pm$45   & 14.6$\pm$2.3 & 2.17 &  570$\pm$50  &  \\ 
  &   & 36$\pm$2 & 0.54$\pm$0.03  &  0.37$\pm$01.5     &  21$\pm$9       & 1530$\pm$40 &  -120$\pm$50 & & &    \\ \hline
0234-07 &  Blue &   32.5$\pm$0.9 & 0.48$\pm$0.02   & 0.50$\pm$0.04     &  11.1$\pm$0.9    &  310$\pm$70   & 0$\pm$50  & 12.6$\pm$0.6  & 2.77&  360$\pm$40 \\ 
  &   &   32$\pm$1  &  0.48$\pm$0.06     &  0.40$\pm$0.06       & 14$\pm$2 & 950$\pm$30 &  -270$\pm$50  &   &   \\ \hline
0955+03  &  Blue &   0.9$\pm$0.2 &   0.13$\pm$0.04 & Unk.    &  Unk.     & 470$\pm$80  & 0$\pm$20 & 16$\pm$4  &   1.54 & 740$\pm$50\\ 
 &   &  3.1$\pm$0.4  &   0.47$\pm$0.08   &  Unk. & Unk.       & 1110$\pm$80  &  190$\pm$50  & \\ 
  &  &   2.6$\pm$0.3  &   0.40$\pm$0.06   & Unk. & Unk.       & 2500$\pm$200  &   -620$\pm$140 &  &   \\ \hline
1153+03 & No & 26$\pm$1    &   0.28$\pm$0.01  &  Unk.  &   Unk. & 270$\pm$10   & 5$\pm$10   &  14.1$\pm$0.1 &15.7 & 360$\pm$20\\ 
  &  &   58$\pm$1 &   0.62$\pm$0.02   &   Unk.   &  14$\pm$4  & 730$\pm$40  & -5$\pm$10  \\ 
  &  &  9$\pm$2  &  0.10$\pm$0.01    & Unk. &  Unk. &  1450$\pm$80  & 330$\pm$70\\  \hline
1228+00  &  Red & 6.4$\pm$0.3 & 0.19$\pm$0.01   & Unk.     & Unk.      & 550$\pm$20   & -60$\pm$10   &12$\pm$1 &   7.36  & 890$\pm$80 \\ 
  &   &    4.6$\pm$0.5 & 0.14$\pm$0.2     &  Unk.      &  Unk.  & 780$\pm$60  & 640$\pm$30 \\ 
  &   &    22.2$\pm$0.4 &     0.67$\pm$0.02 &    Unk.    & Unk. & 2300$\pm$70 & 770$\pm$20  \\ \hline
1307-02  & Blue &     17$\pm$2 & 0.34$\pm$0.04   &  0.23$\pm$0.03   &    16$\pm$2  & 250$\pm$10  & -13$\pm$5  & 10.9$\pm$0.3 & 3.21  & 290$\pm$8\\ 
 &   &    19$\pm$2 & 0.40$\pm$0.04     &   0.61$\pm$0.05     & 7.0$\pm$0.8 & 480$\pm$20   & 56$\pm$9 \\ 
  &  &    13.6$\pm$0.7 &  0.26$\pm$0.02    &   0.16$\pm$0.04     & 18$\pm$4 & 1000$\pm$30   & -190$\pm$20   &  & \\  \hline
1337-01   &  Blue &  9.5$\pm$0.4 &   0.30$\pm$0.01     &  Unk.  &  Unk.    & 350$\pm$20 & 30$\pm$6   &  25$\pm$2*   & 2.03  &  440$\pm$40 \\ 
        & &   22.3$\pm$0.5 & 0.70$\pm$0.02     &  Unk. &   Unk.   &    1280$\pm$60 &  -290$\pm$10 &  &  \\  \hline
 1407+02 &  Blue &  55$\pm$1 &   0.50$\pm$0.02  &  0.57$\pm$0.03     &   9.4$\pm$0.5 & 290$\pm$10  &  10$\pm$3 &  10.9$\pm$0.4 & 2.03  & 320$\pm$8 \\ 
  &   &     41$\pm$2 &     0.37$\pm$0.02     &  0.43$\pm$0.04  &  9.0$\pm$0.9 & 720$\pm$20 & -100$\pm$10 \\ 
  &   &  15.0$\pm$0.6  &   0.13$\pm$0.01   &  $<$0.18   &  $>$6  &  1670$\pm$120 &  -70$\pm$30 \\       \hline
 1413-01 &  No & 14.2$\pm$2 &  0.20$\pm$0.02  &   0.19$\pm$0.02  &   9$\pm$1   &   260$\pm$20  & -20$\pm$10  & 9.6$\pm$0.4  &  3.07 & 400$\pm$8 \\ 
 &   &   45$\pm$2  &  0.61$\pm$0.03   &    0.56$\pm$0.04    &  10.0$\pm$0.6 & 600$\pm$20 & 40$\pm$8  \\ 
  &  &    14$\pm$2  &  0.19$\pm$0.02    &   0.25$\pm$0.03     &  7$\pm$1 &  1190$\pm$50   & -145$\pm$30 \\ \hline
1546-01 & Blue &     7.6$\pm$0.1& 0.58$\pm$0.02   &   0.71$\pm$0.08  &    6.5$\pm$0.7  & 340$\pm$20  & 00$\pm$20 &    8.3$\pm$0.3   & 5.21 & 350$\pm$20\\ 
  &   &    5.5$\pm$0.5  &  0.42$\pm$0.04    &   0.29$\pm$0.08     &   12$\pm$3.5    &   780$\pm$30  & -270$\pm$35   &  \\ \hline
\end{tabular}
\caption{Results of the  spectral decomposition analysis of [OIII] and H$\beta$. (1): Object (shortened) name.
(2): Sign of the [OIII] profile asymmetry implied by AI10 (Table 1). (3) Flux in units of $\times$10$^{-16}$ erg s$^{-1}$ cm$^{-2}$. (4) Fractional contribution of a given kinematic component to the total  [OIII] flux. (5) Idem for H$\beta$. (6) Ionization level as measured from the [OIII]/H$\beta$ ratio. (7) FWHM. (8) Velocity shift $V_S$ relative to the  [OIII] narrow core. (9) [OIII]/H$\beta$ calculated with the total line fluxes (rather than the individual kinematic components as in (6)). (10) Total [OIII] luminosity in units of 10$^{42}$ erg s$^{-1}$. (11) Stellar velocity dispersion  calculated according to  FWHM[OIII]$_{core}$=(1.24$\pm$0.76)$\times$FWHM$_{stars}$  (Greene \& Ho 2005).  The errors do not account for the scatter of the correlation  (see $\delta$3.2).}
\end{table*}

\section{Discussion}

\subsection{Outflows in the NLR of type 2 quasars.}

We have found  evidence for  outflows of ionized gas in the central regions ($r\la$few kpc) of most type 2 quasars in our sample (11/13).  Six  objects in the sample were selected on the basis (among other things) of prior evidence of nuclear kinematic perturbation (Villar-Mart\'\i n et al. 2011) such as  broad and/or highly asymmetric [OIII] lines. However, not obvious bias was applied to give preference to blue  vs. red asymmetries. Moreover, six out of the seven objects for which kinematic criteria were not applied also contain  outflows. So, in spite of the apparent selection bias, the results suggest that the outflow phenomenon is common in the NLR of type 2 quasars. This is consistent with \cite{heck84}, who found  similar evidence in the NLR of  the majority
of radio quiet type 1 quasars (see also Leipski \& Bennert \citeyear{leip06}). 

	 The outflowing gas shows   high ionization level with [OIII]/H$\beta$ sometimes $\ga10$, and often similar or larger than the more quiescent gas.   			  A similar result has been found in some radio quiet type 1 quasars (e.g. Leipski \& Bennert \citeyear{leip06}) and Seyfert galaxies (Veilleux \citeyear{vei91}).  Independently of the outflow origin, such high excitation 
			 suggests that stars are not responsible for ionizing the outflowing gas, but rather AGN related phenomena. 			 This supports that the outflows are within the quasar ionization cones, as commented above. Although contribution from the EELR is expected in the spectra of at least some objects, the line emission is likely to be dominated by the  NLR so that the outflows are expected to be located in this region.

The outflowing gas emits very broad lines (FWHM$\ga$2$\times$FWHM$_ {stars}$, with FWHM in the range 780$\pm$30  to 2500$\pm$200 km s$^{-1}$) and  
 {\it projected} shifts  relative to the systemic  velocity in the range  $V_S\sim$10$\pm$40  and  up to $\sim$-620$\pm$140 km s$^{-1}$ relative to the systemic velocity.   
The $V_S$  are  in general     much less extreme than the motions implied by the large FWHM.  Very broad lines can be produced by acceleration of clouds by or behind the bow shock generated by the outflow (e.g. Villar-Mart\'\i n et al. \citeyear{vil99}). Also, it must be taken into account that the deprojected velocity shifts are likely to be larger, since these are type 2 objects and the cones axis is close to the plane of the sky.    
			 
	We have found a anticorrelation such that the strongest the kinematic perturbation (larger  $R$),
	the lowest the relative contribution of the outfowing gas to the total [OIII] line (smaller $R2$).  Since there is no obvious change in the ionization level of the outfowing gas relative to the ambient, non perturbed gas,  this  anticorrelation  possibly reflects that, the more perturbed the gas is, the smaller the gas mass fraction the outflow involves. 
	
	Based on the spatially resolved behaviour of the FWHM and $V_S$ of [OIII] line in a sample of 15 type 2 quasars at 0.11$\le z \le$0.43,
	\cite{green11}  also proposed that outflows of ionized gas are present in their sample.
	
 \subsection{Outflow mass and mass outflow rate}

	 The  mass  of the ionized outflowing gas $M$  can be estimated with:
	
	 $$M = \frac{1.4~m_p~L(H\beta)}{n_e  ~h~\nu_{H\beta}~\alpha^{eff}_{H\beta}} $$

where L(H$\beta)$ is the luminosity of the broad H$\beta$ component, $n_e$ the density of the outflowing gas, $m_p$  the proton mass,
 $\alpha_{H\beta}^{eff}$ the effective H recombination  coefficient for H$\beta$  (Osterbrock 1989) and $h\nu_{H\beta}$ is the energy of an H$\beta$ photon.   
  
 This refers to the outflowing  gas
 detected within the slit. 
If the outflows are constrained to  a region of size $r<$few kpc, we expect $M$ to provide   an adequate measurement of the total outflowing gas, otherwise  (e.g. Humphrey et al.  \citeyear{hum10}) the inferred masses are lower limits.

 Large uncertainties
 affect this calculation since the unreddened H$\beta$ luminosity is unknown and we have no density measurements.  
The spectral fits imply observed  L(H$\beta)$
 in the range (0.2-2.1)$\times$10$^{41}$ erg s$^{-1}$.

 Using the [SII] doublet from the SDSS spectra of type 2 quasars, Greene et al. (\citeyear{green11}) find a range of densities $\sim$250-500 cm$^{-3}$, with a mean value of 335 cm$^{-3}$.    [OII]$\lambda$3727 is within the spectral range for SDSS J0217-00. The outflow is also evident in this line, which presents a clear blue asymmetry.   Given the clear detection of this wing, 
 it is reasonable to think that $n$ is below the critical density of this line $n_{crit}$=3$\times$10$^3$ cm$^{-3}$.   The
  gas densities estimated 
 in other objects (radio galaxies, Seyferts, far infrared galaxies with galactic superwinds) for the outflowing (nuclear or extended) gas cover a broad range from $\sim$several hundred cm$^{-3}$ (e.g Villar-Mart\'\i n et al. 1999, Arribas \& Mediavilla 1993, Heckman, Miley \& Armus \citeyear{heck90}) to $n_e>$10$^{4.5}$ (Axon et al. 1998). $n  =$250 cm$^{-3}$ seems therefore a reasonable lower limit. Thus, we
 assume a range of densities $n\sim$250 -  3000 cm$^{-3}$ for the outflowing gas.  
 The implied values for  $M$  using the uncorrected L(H$\beta$) are in the range  $\sim$(0.3-1.5)$\times$10$^7$ M$_ {\odot}$ for $n  \sim$250 cm$^{-3}$ and depending on the object. The values are 120 times lower for $n  \sim$3000 cm$^{-3}$. \cite{green11} propose $\sim$10$^{7}$ M$_{\odot}$ as a strict lower limit assuming a gas density $\sim$few$\times$100 cm$^{-3}$. It must be taken into account that these
 authors argue that   all the H$\beta$ emitting gas is involved in the outflow (we find this is not the case), and that much lower densties are possible. 
 
 For an illustrative comparison, the masses involved in {\it neutral} and {\it molecular} outflows in Seyfert and ultraluminous infrared galaxies (ULIRGs), which are  constrained  to spatial scales of $<<$few kpc, are  $>$10$^8$ M$_{\odot}$, and often  $>$10$^9$ M$_{\odot}$  (e.g. Rupke, Veilleux \& Sanders \citeyear{rup05}, Sturm et al. \citeyear{sturm11}).

  It is obvious that better constraints on the  intrinsic L(H$\beta$) and density of the outflowing gas are essential to
  obtain more tightly constrained masses. It is also essential to investigate the existence of outflows in other gas phases. 
Indeed,  outflow studies in different types of galaxies suggest that the ionized phase reveals a very small fraction
of the total outflowing mass, which is likely to be dominated by the molecular and neutral phases. 
As an example, the mass of ionized gas in the (galactic) superwind region of the M82 starburst galaxy is 
  $\sim$2$\times$10$^{5}$ M$_{\odot}$ (Heckman, Armus \& Miley \citeyear{heck90}), while the molecular mass involved in the outflow is        $>$3$\times$10$^8$ M$_{\odot}$ (Walter, Weiss \& Scoville \citeyear{walt02}).

	The mass outflow rate ${\dot M}$ can be calculated as (Dopita \& Sutherland \citeyear{dop03}):
	
$${\dot M} = \frac{3 ~v_{o}~ L(H\beta)    m_p}{r~ n~  h \nu _{H\beta} \alpha_{H\beta}^{eff}} = \frac{2.1~v_0~M  }{r} $$
	 	
		 Assuming a spherical geometry,  $v_0$ is the expanding velocity of the outflowing bubble and $r$ its radius.
 		 In addition to  those affecting $M$,  additional
		 uncertainties are involved for	 $v_0$, which is likely to be $>$$V_s$ (Table 2) and
		  $r$. Although different works suggest that the radius of the NLR in quasars is $\sim$1-several kpc (e.g. Bennert et al. \citeyear{benn02}, Greene et al. \citeyear{green11}), this is only an upper limit for $r$, since the
		size of the outflowing region could be smaller. 
		 
		In spite of the uncetainties, it is interesting to study which conditions are required to produce different 
		$\dot M  $.   
		  For instance, mass flow rates of the NLR outflows discovered in  Seyfert and nearby narrow line radio galaxies are
		  typicially $\la$few M$_{\odot}$ yr$^{-1}$ (Rupke, Veilleux \&  Sanders \citeyear{rup05},  Heckman et al. \citeyear{heck81}; see also Humphrey et al. \citeyear{hum10}).  Such values are certainly possible in our sample within a rather broad  range of allowed parameter values.
		  		  
		  More restrictive conditions are required to produce   ${\dot M}$$>$50  M$_{\odot}$ yr$^{-1}$ 	as observed in some galactic scale  outflows powered by active star formation  (e.g. Heckman, Armus \& Miley \citeyear{heck90}; Lehnnert \& Heckman \citeyear{lenh96};  Rupke et al. \citeyear{rup02}). 
Let us consider a  scenario which favours the efficiency of the outflow: low densities   ($n_e  $=250 cm$^{-3}$) and large line luminosity corrections  (e.g. intrinsic L(H$\beta  $)$\sim$10$\times$observed L(H$\beta  $)).  With these favourable prerequisites, the size of the outflowing region must be $r<<$1kpc. 
Studies of ionized gas outflows using UV and X ray absorption lines indicate that such small distances  relative to he continuum source are possible. If the outflow region has $r\sim$several kpc as argued by \cite{green11}, it is very unlikely that such high mass outflow  rates are generated, since most of the outflowing gas should be collimated within a very narrow cone near the plane of the sky for {\it all} objects.

It is clearly  essential to determine the location of the outflows  to constrain the mass outflow rates.

\section{Conclusions}

We have analyzed the spatially integrated kinematic properties of the  NLR gas in a sample of 13 optically selected SDSS type 2 quasars at $z\sim$0.3-0.6 using  the [OIII]$\lambda\lambda$4959,5007 lines. Optical long-slit spectroscopy at
moderate spectral resolution R$\sim$900-1300  obtained with FORS2 on the VLT has been used for this purpose.

The line profiles show a significant asymmetry in 11 of the 13 quasars. There is a clear preference for blue asymmetries, which are found in 9/13 objects. 
An excess of blue line asymmetries have been observed
in other active classes and non active galaxies. They are suggestive of  outflows, such that the receding expanding gas is preferentially obscured. We propose the same interpretation. The  outflows we see are constrained within  spatial scales of $r<$few kpc from the central engine.  

This scenario is favoured by other results we find:  in addition to more quiescent gas, whose kinematic properties are consistent with gravitational motions, we have discovered highly perturbed gas  in all objects. This gas emits very broad lines ($R=\frac{FWHM[OIII]}{FWHM_ {stars}}\ga$2). While the quiescent gas shows small or null velocity shifts relative to the systemic velocity, the highly perturbed gas trends to show larger shifts. These,  moreover, are blueshifts in general.
In each object the most perturbed gas  trends to have the largest blueshift as well.
 All together support that the perturbed gas, which is responsible for the blue asymmetries  on the line profiles, is outflowing. 
 
The {\it projected} velocity shifts relative to the narrow core of the [OIII] line, which we have considered a reasonable tracer of
the systemic velocity,     are in the range  $\sim$10$\pm$40  and  up to $\sim$-620$\pm$140 km s$^{-1}$, with a median value of $\sim$-145 km s$^{-1}$.  The intrinsic expanding velocities are likely to be larger, given the orientation of
the objects. We FWHM for the outflowing gas in the range 780$\pm$30  to 2500$\pm$200 km s$^{-1}$ with a median value of 1280 km s$^{-1}$.  
The relative contribution of the outflowing gas  
	to the total [OIII] line flux varies from object to object in the range $\sim$10-70\%.  An anticorrelation is found such that, the more perturbed the outflowing gas is (larger $R$), the lower its relative contribution is  to the total [OIII] flux . 
Our results suggest that outflows with more perturbed kinematics  involve a smaller fraction of the total
 mass of ionized gas.

The outflowing gas is in general very highly ionized, as suggested by the large [OIII]/H$\beta$ values (often $\ga$10).	
Independently of the outflows origin, this suggests that the gas is ionized by the quasar, rather than stars, an therefore it is located within the quasar ionization cones.

We argue that our results suggest that nuclear
dusty outflows are a common phenomenon in optically selected type 2 quasars at 0.3$\la z\la$0.6.

\section*{Acknowledgments}
This work has been funded with support from the Spanish Ministerio de Ciencia e Innovaci\'on through the grants 
AYA2007-64712, AYA2009-13036-C02-01 and ESP2007-65475-C02-01 and co-financed with FEDER funds.   Thanks to Clive Tadhunter and the staff at Paranal Observatory for their support during the observations.

\end{document}